\documentclass[longauth]{aa}
\usepackage{graphicx}
\usepackage[varg]{txfonts}
\usepackage[colorlinks = true,
            linkcolor = blue,
            urlcolor  = blue,
            citecolor = blue]{hyperref}
\usepackage{euclid}
\usepackage{bm}
\usepackage{xcolor}
\def\lcdm{$\Lambda$CDM}

\newcommand{\Om}{\Omega_\mathrm{m}}
\newcommand{\Ob}{\Omega_\mathrm{b}}
\newcommand{\Ol}{\Omega_\Lambda}

\newcommand{\Ode}{\Omega_\mathrm{de}}
\newcommand{\void}{\mathrm{v}}

\newcommand{\gal}{\mathrm{g}}

\newcommand{\xibar}{\overline{\xi}}
\newcommand{\C}{\tens{C}}
\newcommand{\DA}{D_\mathrm{M}}

\usepackage{ulem}
\usepackage{multirow}
\usepackage{booktabs}
\usepackage{relsize}
\usepackage[nameinlink,noabbrev]{cleveref}
\Crefname{equation}{Eq.}{Eqs.}
\Crefname{eqnarray}{Eq.}{Eqs.}
\Crefname{section}{Sect.}{Sects.}
\Crefname{figure}{Fig.}{Figs.}
\crefname{equation}{Equation}{Equations}
\crefname{section}{Section}{Sections}
\crefname{figure}{Figure}{Figures}

\begin{document}

\title{\Euclid: Forecasts from redshift-space distortions and the Alcock–Paczynski test with cosmic voids\thanks{This paper is published on behalf of the Euclid Consortium.}}

\titlerunning{RSD \& AP with voids in \Euclid}

\author{N.~Hamaus$^{1}$\thanks{\email{n.hamaus@physik.lmu.de}}, M.~Aubert$^{2,3}$, A.~Pisani$^{4}$, S.~Contarini$^{5,6,7}$, G.~Verza$^{8,9}$, M.-C.~Cousinou$^{3}$, S.~Escoffier$^{3}$, A.~Hawken$^{3}$, G.~Lavaux$^{10}$, G.~Pollina$^{1}$
% ...
}

\institute{$^{1}$ Universit\"ats-Sternwarte M\"unchen, Fakult\"at f\"ur Physik, Ludwig-Maximilians-Universit\"at M\"unchen, Scheinerstrasse 1, 81679 M\"unchen, Germany\\
$^{2}$ University of Lyon, UCB Lyon 1, CNRS/IN2P3, IUF, IP2I Lyon, France\\
$^{3}$ Aix-Marseille Univ, CNRS/IN2P3, CPPM, Marseille, France\\
$^{4}$ Department of Astrophysical Sciences, Peyton Hall, Princeton University, Princeton, NJ 08544, USA\\
$^{5}$ Dipartimento di Fisica e Astronomia ''Augusto Righi'' - Alma Mater Studiorum Universit\'a di Bologna, Viale Berti Pichat 6/2, I-40127 Bologna, Italy\\
$^{6}$ INFN-Sezione di Bologna, Viale Berti Pichat 6/2, I-40127 Bologna, Italy\\
$^{7}$ INAF-IASF Bologna, Via Piero Gobetti 101, I-40129 Bologna, Italy\\
$^{8}$ INFN-Padova, Via Marzolo 8, I-35131 Padova, Italy\\
$^{9}$ Dipartimento di Fisica e Astronomia “G.Galilei", Universit\'a di Padova, Via Marzolo 8, I-35131 Padova, Italy\\
$^{10}$ Sorbonne Universit{\'e}s, UPMC Univ Paris 6 et CNRS, UMR 7095, Institut d'Astrophysique de Paris, 98 bis bd Arago, 75014 Paris, France\\
$^{11}$ Max Planck Institute for Extraterrestrial Physics, Giessenbachstr. 1, D-85748 Garching, Germany\\
$^{12}$ INFN-Sezione di Genova, Via Dodecaneso 33, I-16146, Genova, Italy\\
$^{13}$ Dipartimento di Fisica, Universit\'a degli studi di Genova, and INFN-Sezione di Genova, via Dodecaneso 33, I-16146, Genova, Italy\\
$^{14}$ INFN-Sezione di Milano, Via Celoria 16, I-20133 Milano, Italy\\
$^{15}$ INAF-IASF Milano, Via Alfonso Corti 12, I-20133 Milano, Italy\\
$^{16}$ Dipartimento di Fisica "Aldo Pontremoli", Universit\'a degli Studi di Milano, Via Celoria 16, I-20133 Milano, Italy\\
$^{17}$ INAF-Osservatorio Astronomico di Brera, Via Brera 28, I-20122 Milano, Italy\\
$^{18}$ Universidad de la Laguna, E-38206, San Crist\'{o}bal de La Laguna, Tenerife, Spain\\
$^{19}$ Instituto de Astrof\'{i}sica de Canarias, Calle V\'{i}a L\`actea s/n, 38204, San Crist\`obal de la Laguna, Tenerife, Spain\\
$^{20}$ Dipartimento di Fisica e Astronomia “Augusto Righi” - Alma Mater Studiorum Università di Bologna, via Piero Gobetti 93/2, I-40129 Bologna, Italy\\
$^{21}$ INAF-Osservatorio di Astrofisica e Scienza dello Spazio di Bologna, Via Piero Gobetti 93/3, I-40129 Bologna, Italy\\
$^{22}$ Department of Physics and Astronomy, University of Waterloo, Waterloo, Ontario N2L 3G1, Canada\\
$^{23}$ Centre for Astrophysics, University of Waterloo, Waterloo, Ontario N2L 3G1, Canada\\
$^{24}$ Perimeter Institute for Theoretical Physics, Waterloo, Ontario N2L 2Y5, Canada\\
$^{25}$ Institute of Theoretical Astrophysics, University of Oslo, P.O. Box 1029 Blindern, N-0315 Oslo, Norway\\
$^{26}$ Swedish Collegium for Advanced Study, Thunbergsv\"{a}gen 2, SE-752 38 Uppsala, Sweden\\
$^{27}$ Theoretical astrophysics, Department of Physics and Astronomy, Uppsala University, Box 515, SE-751 20 Uppsala, Sweden\\
$^{28}$ Universit\'e St Joseph; UR EGFEM, Faculty of Sciences, Beirut, Lebanon\\
$^{29}$ Institut de Recherche en Astrophysique et Plan\'etologie (IRAP), Universit\'e de Toulouse, CNRS, UPS, CNES, 14 Av. Edouard Belin, F-31400 Toulouse, France\\
$^{30}$ Mullard Space Science Laboratory, University College London, Holmbury St Mary, Dorking, Surrey RH5 6NT, UK\\
$^{31}$ INAF-Osservatorio Astrofisico di Torino, Via Osservatorio 20, I-10025 Pino Torinese (TO), Italy\\
$^{32}$ INFN-Sezione di Roma Tre, Via della Vasca Navale 84, I-00146, Roma, Italy\\
$^{33}$ Department of Mathematics and Physics, Roma Tre University, Via della Vasca Navale 84, I-00146 Rome, Italy\\
$^{34}$ INAF-Osservatorio Astronomico di Capodimonte, Via Moiariello 16, I-80131 Napoli, Italy\\
$^{35}$ Instituto de Astrof\'isica e Ci\^encias do Espa\c{c}o, Universidade do Porto, CAUP, Rua das Estrelas, PT4150-762 Porto, Portugal\\
$^{36}$ Port d'Informaci\'{o} Cient\'{i}fica, Campus UAB, C. Albareda s/n, 08193 Bellaterra (Barcelona), Spain\\
$^{37}$ Institut de F\'{i}sica d’Altes Energies (IFAE), The Barcelona Institute of Science and Technology, Campus UAB, 08193 Bellaterra (Barcelona), Spain\\
$^{38}$ Institute of Space Sciences (ICE, CSIC), Campus UAB, Carrer de Can Magrans, s/n, 08193 Barcelona, Spain\\
$^{39}$ Institut d’Estudis Espacials de Catalunya (IEEC), Carrer Gran Capit\'a 2-4, 08034 Barcelona, Spain\\
$^{40}$ INAF-Osservatorio Astronomico di Roma, Via Frascati 33, I-00078 Monteporzio Catone, Italy\\
$^{41}$ Department of Physics "E. Pancini", University Federico II, Via Cinthia 6, I-80126, Napoli, Italy\\
$^{42}$ INFN section of Naples, Via Cinthia 6, I-80126, Napoli, Italy\\
$^{43}$ INAF-Osservatorio Astrofisico di Arcetri, Largo E. Fermi 5, I-50125, Firenze, Italy\\
$^{44}$ Institut national de physique nucl\'eaire et de physique des particules, 3 rue Michel-Ange, 75794 Paris C\'edex 16, France\\
$^{45}$ Centre National d'Etudes Spatiales, Toulouse, France\\
$^{46}$ Institute for Astronomy, University of Edinburgh, Royal Observatory, Blackford Hill, Edinburgh EH9 3HJ, UK\\
$^{47}$ European Space Agency/ESRIN, Largo Galileo Galilei 1, 00044 Frascati, Roma, Italy\\
$^{48}$ ESAC/ESA, Camino Bajo del Castillo, s/n., Urb. Villafranca del Castillo, 28692 Villanueva de la Ca\~nada, Madrid, Spain\\
$^{49}$ Univ Lyon, Univ Claude Bernard Lyon 1, CNRS/IN2P3, IP2I Lyon, UMR 5822, F-69622, Villeurbanne, France\\
$^{50}$ Departamento de F\'isica, Faculdade de Ci\^encias, Universidade de Lisboa, Edif\'icio C8, Campo Grande, PT1749-016 Lisboa, Portugal\\
$^{51}$ Instituto de Astrof\'isica e Ci\^encias do Espa\c{c}o, Faculdade de Ci\^encias, Universidade de Lisboa, Campo Grande, PT-1749-016 Lisboa, Portugal\\
$^{52}$ Department of Astronomy, University of Geneva, ch. d\'Ecogia 16, CH-1290 Versoix, Switzerland\\
$^{53}$ Universit\'e Paris-Saclay, CNRS, Institut d'astrophysique spatiale, 91405, Orsay, France\\
$^{54}$ Department of Physics, Oxford University, Keble Road, Oxford OX1 3RH, UK\\
$^{55}$ INAF-Osservatorio Astronomico di Trieste, Via G. B. Tiepolo 11, I-34131 Trieste, Italy\\
$^{56}$ Istituto Nazionale di Astrofisica (INAF) - Osservatorio di Astrofisica e Scienza dello Spazio (OAS), Via Gobetti 93/3, I-40127 Bologna, Italy\\
$^{57}$ Istituto Nazionale di Fisica Nucleare, Sezione di Bologna, Via Irnerio 46, I-40126 Bologna, Italy\\
$^{58}$ INAF-Osservatorio Astronomico di Padova, Via dell'Osservatorio 5, I-35122 Padova, Italy\\
$^{59}$ Jet Propulsion Laboratory, California Institute of Technology, 4800 Oak Grove Drive, Pasadena, CA, 91109, USA\\
$^{60}$ von Hoerner \& Sulger GmbH, Schlo{\ss}Platz 8, D-68723 Schwetzingen, Germany\\
$^{61}$ Max-Planck-Institut f\"ur Astronomie, K\"onigstuhl 17, D-69117 Heidelberg, Germany\\
$^{62}$ AIM, CEA, CNRS, Universit\'{e} Paris-Saclay, Universit\'{e} de Paris, F-91191 Gif-sur-Yvette, France\\
$^{63}$ Universit\'e de Gen\`eve, D\'epartement de Physique Th\'eorique and Centre for Astroparticle Physics, 24 quai Ernest-Ansermet, CH-1211 Gen\`eve 4, Switzerland\\
$^{64}$ Department of Physics and Helsinki Institute of Physics, Gustaf H\"allstr\"omin katu 2, 00014 University of Helsinki, Finland\\
$^{65}$ NOVA optical infrared instrumentation group at ASTRON, Oude Hoogeveensedijk 4, 7991PD, Dwingeloo, The Netherlands\\
$^{66}$ Argelander-Institut f\"ur Astronomie, Universit\"at Bonn, Auf dem H\"ugel 71, 53121 Bonn, Germany\\
$^{67}$ Institute for Computational Cosmology, Department of Physics, Durham University, South Road, Durham, DH1 3LE, UK\\
$^{68}$ Universit\'e C\^{o}te d'Azur, Observatoire de la C\^{o}te d'Azur, CNRS, Laboratoire Lagrange, Bd de l'Observatoire, CS 34229, 06304 Nice cedex 4, France\\
$^{69}$ University of Applied Sciences and Arts of Northwestern Switzerland, School of Engineering, 5210 Windisch, Switzerland\\
$^{70}$ California institute of Technology, 1200 E California Blvd, Pasadena, CA 91125, USA\\
$^{71}$ Institute of Physics, Laboratory of Astrophysics, Ecole Polytechnique F\'{e}d\'{e}rale de Lausanne (EPFL), Observatoire de Sauverny, 1290 Versoix, Switzerland\\
$^{72}$ European Space Agency/ESTEC, Keplerlaan 1, 2201 AZ Noordwijk, The Netherlands\\
$^{73}$ Department of Physics and Astronomy, University of Aarhus, Ny Munkegade 120, DK–8000 Aarhus C, Denmark\\
$^{74}$ Institute of Space Science, Bucharest, Ro-077125, Romania\\
$^{75}$ Departamento de Astrof\'{i}sica, Universidad de La Laguna, E-38206, La Laguna, Tenerife, Spain\\
$^{76}$ Centro de Investigaciones Energ\'eticas, Medioambientales y Tecnol\'ogicas (CIEMAT), Avenida Complutense 40, 28040 Madrid, Spain\\
$^{77}$ Instituto de Astrof\'isica e Ci\^encias do Espa\c{c}o, Faculdade de Ci\^encias, Universidade de Lisboa, Tapada da Ajuda, PT-1349-018 Lisboa, Portugal\\
$^{78}$ Universidad Polit\'ecnica de Cartagena, Departamento de Electr\'onica y Tecnolog\'ia de Computadoras, 30202 Cartagena, Spain\\
$^{79}$ Kapteyn Astronomical Institute, University of Groningen, PO Box 800, 9700 AV Groningen, The Netherlands\\
$^{80}$ Infrared Processing and Analysis Center, California Institute of Technology, Pasadena, CA 91125, USA\\
$^{81}$ Dipartimento di Fisica e Astronomia, Universit\'a di Bologna, Via Gobetti 93/2, I-40129 Bologna, Italy\\
$^{82}$ INFN-Sezione di Torino, Via P. Giuria 1, I-10125 Torino, Italy\\
$^{83}$ Dipartimento di Fisica, Universit\'a degli Studi di Torino, Via P. Giuria 1, I-10125 Torino, Italy\\
$^{84}$ Universit\'e de Paris, CNRS, Astroparticule et Cosmologie, F-75013 Paris, France
}

\date{Received 23 August 2021 / Accepted 31 October 2021}
\authorrunning{N. Hamaus \& M. Aubert et al.}

\abstract{
\Euclid is poised to survey galaxies across a cosmological volume of unprecedented size, providing observations of more than a billion objects distributed over a third of the full sky. Approximately 20 million of these galaxies will have their spectroscopy available, allowing us to map the three-dimensional large-scale structure of the Universe in great detail. This paper investigates prospects for the detection of cosmic voids therein and the unique benefit they provide for cosmological studies. In particular, we study the imprints of dynamic (redshift-space) and geometric (Alcock--Paczynski) distortions of average void shapes and their constraining power on the growth of structure and cosmological distance ratios. To this end, we made use of the Flagship mock catalog, a state-of-the-art simulation of the data expected to be observed with \Euclid. We arranged the data into four adjacent redshift bins, each of which contains about~$11\,000$ voids and we estimated the stacked void-galaxy cross-correlation function in every bin. Fitting a linear-theory model to the data, we obtained constraints on~$f/b$ and~$\DA H$, where~$f$ is the linear growth rate of density fluctuations, $b$ the galaxy bias, $\DA$ the comoving angular diameter distance, and~$H$ the Hubble rate. In addition, we marginalized over two nuisance parameters included in our model to account for unknown systematic effects in the analysis. With this approach, \Euclid will be able to reach a relative precision of about~$4\%$ on measurements of~$f/b$ and~$0.5\%$ on~$\DA H$ in each redshift bin. Better modeling or calibration of the nuisance parameters may further increase this precision to~$1\%$ and~$0.4\%$, respectively. Our results show that the exploitation of cosmic voids in \Euclid will provide competitive constraints on cosmology even as a stand-alone probe. For example, the equation-of-state parameter,~$w$, for dark energy will be measured with a precision of about~$10\%$, consistent with previous more approximate forecasts.
}


\maketitle

\section{Introduction}\label{sec:intro}
The formation of cosmic voids in the large-scale structure of the Universe is a consequence of the gravitational interaction of its initially smooth distribution of matter eventually evolving into collapsed structures that make up the cosmic web~\citep{Zeldovich1970}. This process leaves behind vast regions of nearly empty space that constitute the largest known structures in the Universe. Since their first discovery in the late $1970$s~\citep{Gregory1978,Joeveer1978}, cosmic voids have intrigued scientists given their peculiar nature~\citep[e.g.,][]{Kirshner1981,Bertschinger1985,White1987,vdWeygaert1993,Peebles2001}. However, only the recent advances in surveys, such as 6dFGS~\citep{Jones2004}, BOSS~\citep{Dawson2013}, DES~\citep{DES2005}, eBOSS~\citep{Dawson2016}, KiDS~\citep{deJong2013}, SDSS~\citep{Eisenstein2011}, and VIPERS~\citep{Guzzo2014}, have enabled systematic studies of statistically significant sample sizes~\citep[e.g.,][]{Pan2012,Sutter2012a}, placing the long-overlooked field of cosmic voids into a new focus of interest in astronomy. In conjunction with the  extensive development of large simulations~\citep[e.g.,][]{Springel2005,Schaye2015,Dolag2016,Potter2017}, this has sparked a plethora of studies on voids and their connection to galaxy formation~\citep[][]{Hoyle2005,Patiri2006,Kreckel2012,Ricciardelli2014,Habouzit2020,Panchal2020}, large-scale structure~\citep[][]{Sheth2004,Hahn2007,vdWeygaert2009,Jennings2013,Hamaus2014a,Chan2014,Voivodic2020}, the nature of gravity~\citep[][]{Clampitt2013,Zivick2015,Cai2015,Barreira2015,Achitouv2016,Voivodic2017,Falck2018,Sahlen2018,Baker2018,Paillas2019,Davies2019,Perico2019,Alam2021,Contarini2021a,Wilson2020}, properties of dark matter~\citep[][]{Leclercq2015,Yang2015,Reed2015,Baldi2018}, dark energy~\citep[][]{Lee2009,Bos2012,Spolyar2013,Pisani2015a,Pollina2016,Verza2019}, massive neutrinos~\citep[][]{Massara2015,Banerjee2016,Sahlen2019,Kreisch2019,Schuster2019,Zhang2020,Bayer2021}, inflation~\citep[][]{Chan2019}, and cosmology in general~\citep[][]{Lavaux2012,Sutter2012b,Hamaus2014c,Hamaus2016,Correa2019,Correa2022,Contarini2019,Nadathur2019,Nadathur2020,Hamaus2020,Paillas2021,Kreisch2021}. We refer to \cite{Pisani2019} for a more extensive recent summary.

From an observational perspective, voids are an abundant structure type that, together with halos, filaments, and walls, build up the cosmic web. It is therefore natural to utilize them in the search for those observables that have traditionally been measured via galaxies or galaxy clusters, which trace the overdense regions of large-scale structure. This strategy has proven itself to be very promising in recent years, uncovering a treasure trove of untapped signals that carry cosmologically relevant information, such as the integrated Sachs--Wolfe (ISW)~\citep{Granett2008,Cai2010,Cai2014,Ilic2013,Nadathur2016,Kovacs2016,Kovacs2019}, Sunyaev--Zeldovich (SZ)~\citep{Alonso2018}, and Alcock--Paczynski (AP) effects~\citep{Sutter2012b,Sutter2014b,Hamaus2016,Mao2017b,Nadathur2019}, as well as gravitational lensing~\citep{Melchior2014,Clampitt2015,Gruen2016,SanchezC2017,Cai2017a,Chantavat2017,Brouwer2018,Fang2019}, baryon acoustic oscillations (BAO)~\citep{Kitaura2016b,Liang2016,Chan2021,Forero2021}, and redshift-space distortions (RSD)~\citep{Paz2013,Cai2016,Hamaus2017,Hawken2017,Hawken2020,Achitouv2019,Aubert2020}. The aim of this paper is to forecast the constraining power on cosmological parameters with a combined analysis of RSD and the AP effect from voids available in \Euclid. As proposed in~\citet{Hamaus2015} and carried out with BOSS data in~\citet{Hamaus2016} for the first time, this combined approach allows simultaneous constraints on the expansion history of the Universe and the growth rate of structure inside it.

The \Euclid satellite mission is a ``Stage-IV'' dark energy experiment~\citep{Albrecht2006} that will outperform current surveys in the number of observed galaxies and in coverage of cosmological volume~\citep{Laureijs2011}. Scheduled for launch in 2022, an assessment of its science performance is timely~\citep{Amendola2018,Euclid2020} and the scientific return that can be expected from voids is being investigated in a series of companion papers within the Euclid Collaboration. These papers cover different observables, such as the void size function~(\textcolor{blue}{Contarini et al. in prep.}), the void-galaxy cross-correlation function after velocity-field reconstruction~(\textcolor{blue}{Radinovi\'c et al. in prep.}), or void lensing~(\textcolor{blue}{Bonici et al. in prep.}), providing independent forecasts on their cosmological constraining power. In this paper, we present a mock-data analysis of the stacked void-galaxy cross-correlation function in redshift space based on the \textit{Flagship} simulation~\citep{Potter2017}, which provides realistic galaxy catalogs as expected to be observed with \Euclid. In the following, we outline the theoretical background in \Cref{sec:theory}, describe the mock data in \Cref{sec:data}, and present our results in \Cref{sec:analysis}. The implications of our findings are then discussed in \Cref{sec:discussion} and our conclusions are summarized in \Cref{sec:conclusion}.

\section{Theoretical background}\label{sec:theory}
According to the cosmological principle, the Universe obeys homogeneity and isotropy on very large scales, which is supported by recent observations~\citep[e.g.,][]{Scrimgeour2012,Laurent2016,Ntelis2017,Goncalves2021}. However, below the order of $10^2\,\mathrm{Mpc}$ scales, we observe the structures that form the cosmic web, which break these symmetries locally. Nevertheless, the principle is still valid on those scales in a statistical sense, that is, for an ensemble average over patches of similar extent from different locations in the Universe. If the physical size of such patches is known (a so-called standard ruler), this enables an inertial observer to determine cosmological distances and the expansion history. The BAO feature in the galaxy distribution is a famous example for a standard ruler, it has been exploited for distance measurements with great success in the past~\citep[e.g.,][]{Alam2017,Alam2020} and constitutes one of the main cosmological probes of \Euclid~\citep{Laureijs2011}.

A related approach may be pursued with so-called standard spheres, namely, patches of a known physical shape (in particular, spherically symmetric ones). This method has originally been proposed by~\citet{Alcock1979} (AP) as a probe of the expansion history, it was later demonstrated that voids are well suited for this type of experiment~\citep{Ryden1995,Lavaux2012}. In principle, it applies to any type of structure in the Universe that exhibits random orientations (such as halos, filaments, or walls), which necessarily results in a spherically symmetric ensemble average. However, the expansion history can only be probed with structures that have not decoupled from the Hubble flow via gravitational collapse. Furthermore, spherical symmetry is broken by peculiar line-of-sight motions of the observed objects that make up this structure. These cause a Doppler shift in the received spectrum and, hence, affect the redshift-distance relation to the source~\citep{Kaiser1987}. In order to apply the AP test, one has to account for those RSD, which requires the modeling of peculiar velocities. For the complex phase-space structure of halos, respectively galaxy clusters as their observational counterparts, this is a very challenging problem. For filaments and walls, the situation is only marginally improved, since they have experienced shell crossing in at least one dimension. On the other hand, voids have hardly undergone any shell crossing in their interiors~\citep{Shandarin2011,Abel2012,Sutter2014c,Hahn2015}, providing an environment that is characterized by a coherent flow of matter and on that is therefore very amenable to dynamical models.

In fact, with the help of $N$-body simulations, it has been shown that local mass conservation provides a very accurate description, even at linear order in the density fluctuations~\citep{Hamaus2014b}. In that case, the velocity field $\vec{u}$ relative to the void center is given by~\citep{Peebles1980}
\begin{equation}
        \vec{u}(\vec{r}) = -\frac{f(z)}{3}\frac{H(z)}{1+z}\,\Delta(r)\,\vec{r}\;,
        \label{u(r)}
\end{equation}
where $\vec{r}$ is the comoving real-space distance vector to the void center, $H(z)$ the Hubble rate at redshift $z$, $f(z)$ is the linear growth rate of density perturbations $\delta$, and $\Delta(r)$ is the average matter-density contrast enclosed in a spherical region of radius~$r$:
\begin{equation}
        \Delta(r) = \frac{3}{r^3}\int_0^r\delta(r')\,r'^{\,2}\,\mathrm{d}r'\;. \label{Delta(r)}
\end{equation}
The comoving distance vector $\vec{s}$ in redshift space receives an additional contribution from the line-of-sight component of $\vec{u}$ (indicated by $\vec{u}_\parallel$), caused by the Doppler effect,
\begin{equation}
        \vec{s} = \vec{r} + \frac{1+z}{H(z)}\,\vec{u}_\parallel = \vec{r} - \frac{f(z)}{3}\,\Delta(r)\,\vec{r}_\parallel \;.
        \label{s(r)}
\end{equation}
This equation determines the mapping between real and redshift space at linear order. Its Jacobian can be expressed analytically and yields a relation between the void-galaxy cross-correlation functions $\xi$ in both spaces (a superscript $s$ indicates redshift space),
\begin{equation}
        \xi^s(\vec{s}) \simeq \xi(r) + \frac{f}{3}\Delta(r) + f\mu^2\brackets{\delta(r)-\Delta(r)}\;,
        \label{xi^s_lin}
\end{equation}
where $\mu=r_\parallel/r$ denotes the cosine of the angle between $\vec{r}$ and the line of sight~\citep[see][for a more detailed derivation]{Cai2016,Hamaus2017,Hamaus2020}. The real-space quantities $\xi(r)$, $\delta(r)$ and its integral $\Delta(r)$ on the right-hand side of \Cref{xi^s_lin} are a priori unknown, but they can be related to the observables with some basic assumptions. Firstly, $\xi(r)$ can be obtained via deprojection of the projected void-galaxy cross-correlation function $\xi^s_\mathrm{p}(s_\perp)$ in redshift space~\citep{Pisani2014,Hawken2017},
\begin{equation}
        \xi(r) = -\frac{1}{\pi}\int_r^\infty\frac{\mathrm{d}\xi^s_\mathrm{p}(s_\perp)}{\mathrm{d}s_\perp}\left(s_\perp^2-r^2\right)^{-1/2}\mathrm{d}s_\perp\;.
        \label{xi_d}
\end{equation}
By construction $\xi^s_\mathrm{p}(s_\perp)$ is insensitive to RSD, since the line-of-sight component $s_\parallel$ is integrated out in its definition and the projected separation $s_\perp$ on the plane of the sky is identical to its real-space counterpart $r_\perp$,
\begin{equation}
        \xi^s_\mathrm{p}(s_\perp) = \int\xi^s(\vec{s})\,\mathrm{d}s_\parallel =  2\int_{s_\perp}^\infty r\,\xi(r)\left(r^2-s_\perp^2\right)^{-1/2}\mathrm{d}r\;.
        \label{xi_p}
\end{equation}
\cref{xi_d,xi_p} are also referred to as inverse and forward Abel transform, respectively~\citep{Abel1842,Bracewell1999}.

Secondly, the matter fluctuation $\delta(r)$ around the void center can be related to $\xi(r)$ assuming a bias relation for the galaxies in that region. Based on simulation studies, it has been demonstrated that a linear relation of the form $\xi(r) = b\delta(r)$ with a proportionality constant $b$ serves that purpose with sufficiently high accuracy. Moreover, it has been shown that the value of $b$ is linearly related to the large-scale linear galaxy bias of the tracer distribution, and coincides with it for sufficiently large voids~\citep{Sutter2014a,Pollina2017, Pollina2019,Contarini2019,Ronconi2019}. With this, \Cref{xi^s_lin} can be expressed as
\begin{equation}
        \xi^s(\vec{s}) \simeq \xi(r) + \frac{1}{3}\frac{f}{b}\xibar(r) + \frac{f}{b}\mu^2\brackets{\xi(r)-\xibar(r)}\;,
        \label{xi^s_lin2}
\end{equation}
where
\begin{equation}
\xibar(r) = 3r^{-3}\!\int_0^r\xi(r')\,r'^2\,\mathrm{d}r'\,.
\end{equation}
Now, together with \Cref{s(r)} to relate real and redshift-space coordinates, \Cref{xi^s_lin2} provides a description of the observable void-galaxy cross-correlation function at linear order in perturbation theory.

In order to determine the distance vector $\vec{s}$ for a given void-galaxy pair, it is necessary to convert their observed separation in angle $\delta\vartheta$ and redshift $\delta z$ to comoving distances via
\begin{equation}
        s_\perp = \DA(z)\,\delta\vartheta\;,\qquad s_\parallel = \frac{c}{H(z)}\,\delta z\;, \label{comoving}
\end{equation}
where $\DA(z)$ is the comoving angular diameter distance. Both $H(z)$ and $\DA(z)$ depend on cosmology, so any evaluation of \Cref{comoving} requires the assumption of a fiducial cosmological model. To maintain the full generality of the model, it is customary to introduce two AP parameters that inherit the dependence on cosmology via~\citep[e.g.,][]{SanchezA2017}
\begin{equation}
        q_\perp = \frac{s_\perp^*}{s_\perp} = \frac{\DA^*(z)}{\DA(z)}\;,\qquad q_\parallel = \frac{s_\parallel^*}{s_\parallel} = \frac{H(z)}{H^*(z)}\;. \label{AP}
\end{equation}
In this notation, the starred quantities are evaluated in the true underlying cosmology, which is unknown, while the un-starred ones correspond to the assumed fiducial values of $\DA$ and $H$. \cref{comoving} can be rewritten as $s_\perp^* = q_\perp\DA(z)\,\delta\vartheta$ and $s_\parallel^* = q_\parallel\,c\,\delta z/H(z)$, which is valid for a wide range of cosmological models. In the special case where the fiducial cosmology coincides with the true one, we have $q_\perp=q_\parallel=1$. This method is known as the AP test, providing cosmological constraints via measurements of $\DA(z)$ and $H(z)$. Without an absolute calibration scale the parameters $q_\perp$ and $q_\parallel$ remain degenerate in the AP test and it is only their ratio, 
\begin{equation}
        \varepsilon = \frac{q_\perp}{q_\parallel} =  \frac{\DA^*(z)H^*(z)}{\DA(z)H(z)}\;,
        \label{epsilon}
\end{equation}
that can be determined, providing a measurement of $\DA^*(z)H^*(z)$. We adopt a flat \lcdm\ cosmology as our fiducial model, where
\begin{equation}
        \DA(z)=\int_0^z\frac{c}{H(z')}\,\mathrm{d}z'\;,\quad
        H(z) = H_0\sqrt{\Om(1+z)^3+\Ol}\;,\label{HDA}
\end{equation}
with the present-day Hubble constant, $H_0$, matter-density parameter, $\Om$, and cosmological constant parameter, $\Ol=1-\Om$. This model includes the true input cosmology of the Flagship simulation with parameter values stated in \Cref{sec:data} below, which is also used for void identification. The impact of the assumed cosmology on the latter has previously been investigated and was found to be negligible~\citep[e.g.,][]{Hamaus2020}. In \Cref{sec:discussion}, we additionally consider an extended $w$CDM model to include the equation-of-state parameter, $w$, for dark energy.

\section{Mock catalogs}\label{sec:data}
\subsection{Flagship simulation}\label{subsec:Flagship}
We employed the \Euclid Flagship mock galaxy catalog (version $\mathrm{1.8.4}$), which is based on an $N$-body simulation of $12\,600^3$ (two trillion) dark matter particles in a periodic box of $3780\,\hMpc$ on a side~\citep{Potter2017}. It adopts a flat \lcdm\ cosmology with parameter values $\Om=0.319$, $\Ob=0.049$, $\Ol=0.681$, $\sigma_8=0.83$, $n_\mathrm{s}=0.96$, and $h=0.67$, as obtained by \Planck in 2015~\citep{Planck2015}. Dark matter halos are identified with the \texttt{ROCKSTAR} halo finder~\citep{Behroozi2013} and populated with central and satellite galaxies using a halo occupation distribution (HOD) framework to reproduce the relevant observables for \Euclid's main cosmological probes. The HOD algorithm~\citep{Carretero2015,Crocce2015} has been calibrated exploiting several observational constraints, including the local luminosity function for the faintest galaxies~\citep{Blanton2003,Blanton2005} and galaxy clustering statistics as a function of luminosity and color~\citep{Zehavi2011}. The simulation box is converted into a light cone that comprises one octant of the sky ($5157$ $\mathrm{deg}^2$). The expected footprint of \Euclid will cover a significantly larger sky area of roughly $15\,000 \ \mathrm{deg}^2$ in total.

With its two complementary instruments, the VISible imager~\citep[VIS,][]{Cropper2018} and the Near Infrared Spectrograph and Photometer~\citep[NISP,][]{Costille2018}, \Euclid will provide photometry and slitless spectroscopy using a near-infrared grism. In this paper, we consider the spectroscopic galaxy sample after a cut in H$\alpha$ flux, $f_{\mathrm{H}\alpha} > 2\times10^{-16} \,\mathrm{erg}\,\mathrm{s}^{-1}\mathrm{cm}^{-2}$, which corresponds to the expected detection threshold for \Euclid, covering a redshift range of $0.9<z<1.8$~\citep{Laureijs2011}. In addition, we randomly dilute this sample to $60\%$ of all objects and add a Gaussian redshift error with RMS of $\sigma_z=0.001$, independent of $z$. This matches the expected median completeness and spectroscopic performance of the survey in a more optimistic scenario and results in a final mock catalog containing about $6.5\expo{6}$ galaxies.

\subsection{VIDE voids}\label{subsec:VIDE} 
For the creation of void catalogs we make use of the public Void IDentification and Examination toolkit \texttt{VIDE}\footnote{\url{https://bitbucket.org/cosmicvoids/vide_public/}}~\citep{Sutter2015}. At the core of \texttt{VIDE} is \texttt{ZOBOV}~\citep{Neyrinck2008}, which is a watershed algorithm~\citep{Platen2007} that delineates three-dimensional basins in the density field of tracer particles. The density field is estimated via Voronoi tessellation, assigning to each tracer particle,~$i$, a cell of volume~$V_i$. In \Euclid, these tracer particles are galaxies, distributed over a masked light cone with a redshift-dependent number density, $n_\mathrm{g}(z)$. In particular, \texttt{VIDE} provides a framework for handling these complications; we refer to \cite{Sutter2015} for a detailed discussion. Local density minima serve as starting points to identify extended watershed basins whose density monotonically increases among their neighboring cells. All the Voronoi cells that make up such a basin determine a void region that includes its volume-weighted barycenter, which we define as the location of the void center.

\begin{figure*}[t]
        \centering
        \resizebox{\hsize}{!}{
                \includegraphics[trim=0 6.5 0 0]{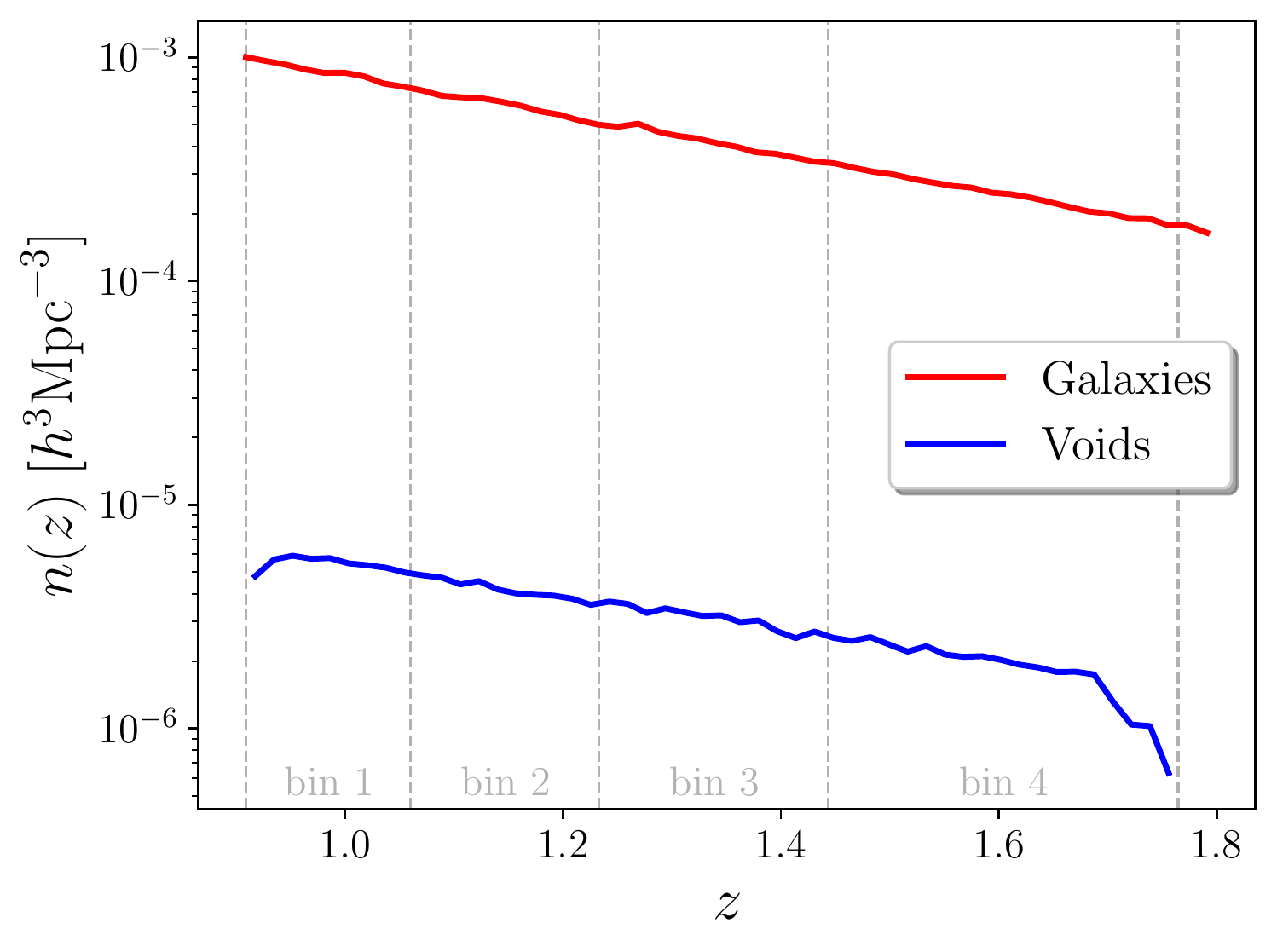}
                \includegraphics[trim=0 10 0 0]{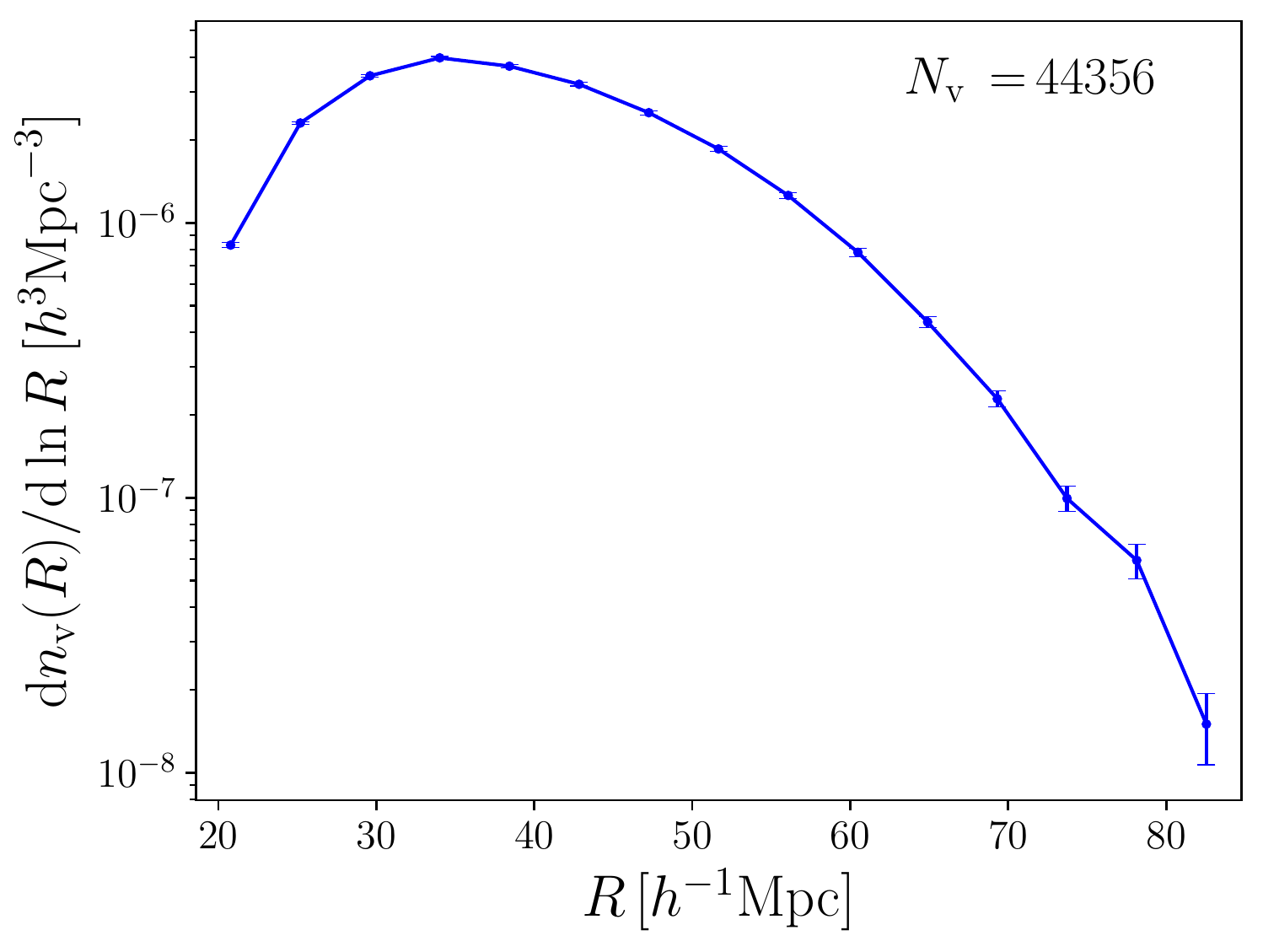}}
        \caption{Properties of tracers and voids in the \Euclid Flagship mock catalog. \textit{Left}: Number density of spectroscopic galaxies and selected \texttt{VIDE} voids as function of redshift; vertical dashed lines indicate the redshift bins. \textit{Right}: Number density of the same \texttt{VIDE} voids as a function their effective radius $R$ (void size function) from the entire redshift range, encompassing a total of $N_\void=44356$ voids with $18.6\,\hMpc<R<84.8\,\hMpc$. Poisson statistics are assumed for the error bars. We refer to our companion paper for a detailed cosmological forecast based on the void size function~(\textcolor{blue}{Contarini et al. in prep.}).}
        \label{fig:nz}
\end{figure*}

Moreover, we assign an effective radius $R$ to every void, which corresponds to a sphere with the same volume:
\begin{equation}
        R = \paren{\frac{3}{4\pi}\sum\nolimits_iV_i}^{1/3}\;.
\end{equation}
Using \texttt{VIDE} we find a total of $N_\void=58\,601$ voids in the Flagship mock light cone, after discarding those that intersect with the survey mask or redshift boundaries. In addition, we implement a purity selection cut based on the effective radius of a void at redshift $Z$,
\begin{equation}
        R > N_\mathrm{s}\paren{\frac{4\pi}{3}n_\mathrm{g}(Z)}^{-1/3}\;.
        \label{mgs}
\end{equation}
We denote the redshift of void centers with a capital $Z$, to distinguish it from the redshift $z$ of galaxies, while $N_\mathrm{s}$ determines the minimum included void size in units of the average tracer separation. The smaller the value of $N_\mathrm{s}$, the larger the contamination by spurious voids that may arise via Poisson fluctuations~\citep{Neyrinck2008,Cousinou2019}  and have been misidentified due to RSD~\citep{Pisani2015b,Correa2021,Correa2022}. We adopt a value of $N_\mathrm{s}=3$, which leaves us with a final number of $N_\void=44\,356$ voids with minimum effective radius of $R\simeq18.6\,\hMpc$. This sample is further split into $4$ consecutive redshift bins with an equal number of voids per bin, $N_\void=11\,089$, see~\Cref{fig:nz}. The selected number of redshift bins is a trade-off between the necessary statistical power to estimate our data vectors and their covariance with sufficient accuracy in each bin, and an adequate sampling of the redshift evolution of $f\sigma_8$ and $\DA H$ (see \Cref{sec:analysis}). The removal of voids close to the redshift boundaries of the Flagship light cone causes their abundance to decline, which lowers the statistical constraining power in that regime.

\section{Data analysis}\label{sec:analysis}
Our data vector is represented by the void-galaxy cross-correlation function in redshift space. As this function is an-isotropic, we can either consider its 2D version with coordinates perpendicular to and along the line of sight, $\xi^s(s_\perp,s_\parallel)$, or its decomposition into multipoles by use of Legendre polynomials $P_\ell$ of order $\ell$,
\begin{equation}
        \xi^s_\ell(s) = \frac{2\ell+1}{2}\int\limits_{-1}^1\xi^s(s,\mu_s)P_\ell(\mu_s)\,\mathrm{d}\mu_s\;,
        \label{Legendre}
\end{equation}
where $\mu_s=s_\parallel/s$. We highlight that the notations $\xi^s(\vec{s})$, $\xi^s(s_\perp,s_\parallel)$, and $\xi^s(s,\mu_s)$ all refer to the same physical quantity, albeit via their different mathematical formulations. Here, we make use of the full 2D correlation function for our model fits, since it contains all the available information on RSD and AP distortions. For the sake of completeness, we additionally provide the three multipoles of the lowest even order, $\ell=0,2,4$ (i.e., monopole, quadrupole, and hexadecapole). Their theoretical linear predictions directly follow from~\Cref{xi^s_lin2}:
\begin{equation}
        \xi^s_0(s) = \paren{1+\frac{f/b}{3}}\xi(r),\;\;
        \xi^s_2(s) = \frac{2f/b}{3}\brackets{\xi(r)-\xibar(r)},\;\;
        \xi^s_4(s) = 0\;.
        \label{multipoles}
\end{equation}

\subsection{Estimator}\label{subsec:analysis}
For our mock measurements of $\xi^s(s_\perp,s_\parallel)$, we utilize the~\cite{Landy1993} estimator for cross correlations,
\begin{equation}
        \hat{\xi}^s(s_\perp,s_\parallel) = \frac{\ave{\mathcal{D}_\void \mathcal{D}_\gal}-\ave{\mathcal{D}_\void \mathcal{R}_\gal}-\ave{\mathcal{R}_\void \mathcal{D}_\gal}+\ave{\mathcal{R}_\void \mathcal{R}_\gal}}{\ave{\mathcal{R}_\void \mathcal{R}_\gal}}\;,
        \label{LS_estimator}
\end{equation}
where the angled brackets signify normalized pair counts of void-center and galaxy positions in the data ($\mathcal{D}_\void$, $\mathcal{D}_\gal$) and corresponding random positions ($\mathcal{R}_\void$, $\mathcal{R}_\gal$) in bins of $s_\perp$ and $s_\parallel$. We chose a fixed binning in units of the effective void radius for each individual void and express all distances in units of $R$ as well. This allows one to coherently capture the characteristic topology of voids from a range of sizes including their boundaries in an ensemble-average sense. The resulting statistic is also referred to as a void stack or stacked void-galaxy cross-correlation function. We have generated the randoms via sampling from the redshift distributions of galaxies and voids as depicted in~\Cref{fig:nz}, but with ten times the number of objects and without spatial clustering. We applied the same angular footprint as for the mock data and additionally assign an effective radius to every void random, drawn from the radius distribution of galaxy voids. The latter is used to express distances from void randoms in units of $R$, consistent with the stacking procedure of galaxy voids. The uncertainty of the estimated $\hat{\xi}^s(s_\perp,s_\parallel)$ is quantified by its covariance matrix:
\begin{equation}
        \hat{\C}_{ij} = \ave{\paren{\hat{\xi}^s(\vec{s}_i)-\langle\hat{\xi}^s(\vec{s}_i)\rangle}\paren{\hat{\xi}^s(\vec{s}_j)-\langle\hat{\xi}^s(\vec{s}_j)\rangle}}\;,
        \label{covariance}
\end{equation}
where angled brackets imply averaging over an ensemble of observations. The square root of the diagonal elements, $\hat{\C}_{ii}$, are used as error bars on our measurements of~$\hat{\xi}^s$. Although we can only observe one universe (respectively a single Flagship mock catalog), ergodicity allows us to estimate $\hat{\C}_{ij}$ via spatial averaging over distinct patches. This naturally motivates the jackknife technique to be applied on the available sample of voids, which are non-overlapping. Therefore, we simply remove one void at a time in the estimator of $\xi^s$ from~\Cref{LS_estimator}, which provides $N_\void$ jackknife samples. This approach has been tested on simulations and validated on mocks in previous analyses~\citep{Paz2013,Cai2016,Correa2019,Hamaus2020}. It has further been shown that, in the limit of large sample sizes, the jackknife technique provides consistent covariance estimates compared to the ones obtained from many independent mock catalogs~\citep{Favole2021}. Residual differences between the two methods indicate the jackknife approach to yield somewhat higher covariances, which renders our error forecast conservative.

\subsection{Model and likelihood}\label{subsec:model}
As previously demonstrated in~\cite{Hamaus2020}, we include two additional nuisance parameters, $\mathcal{M}$ and $\mathcal{Q}$, in the theory model of~\Cref{xi^s_lin2}, enabling us to account for systematic effects. Here, $\mathcal{M}$ (monopole-like) is used as a free amplitude of the deprojected correlation function $\xi(r)$ in real space and $\mathcal{Q}$ (quadrupole-like) is a free amplitude for the quadrupole term proportional to $\mu^2$. Here, we adopt a slightly modified, empirically motivated parametrization of this model, with enhanced coefficients for the Jacobian (second and third) terms in~\Cref{xi^s_lin2},
\begin{equation}
        \xi^s(s_\perp,s_\parallel) = \mathcal{M}\curly{\xi(r) + \frac{f}{b}\xibar(r) + 2\mathcal{Q}\,\frac{f}{b}\mu^2\brackets{\xi(r)-\xibar(r)}}\;.
        \label{xi^s_lin3}
\end{equation}
The parameter $\mathcal{M}$ adjusts for potential inaccuracies arising in the deprojection technique and a contamination of the void sample by spurious Poisson fluctuations, which can attenuate the amplitude of the monopole and quadrupole~\citep{Cousinou2019}. The parameter $\mathcal{Q}$ accounts for possible selection effects when voids are identified in anisotropic redshift space~\citep{Pisani2015b,Correa2021,Correa2022}. For example, the occurrence of shell crossing and virialization affects the topology of void boundaries in redshift space~\citep{Hahn2015}, resulting in the well-known finger-of-God (FoG) effect~\citep{Jackson1972}. In turn, this can enhance the Jacobian terms in~\Cref{xi^s_lin2}, which motivates the empirically determined modification of their coefficients in~\Cref{xi^s_lin3}. A similar result can be achieved by enhancing the values of $\mathcal{M}$ and $\mathcal{Q}$, but keeping the original form of~\Cref{xi^s_lin2}, which can be approximately understood as a redefinition of the nuisance parameters. However, the form of~\Cref{xi^s_lin3} is found to better describe the void-galaxy cross-correlation function in terms of goodness of fit, while at the same time yields nuisance parameters that are distributed more closely around values of one. This approach is akin to other empirical model extensions that have been proposed in the literature~\citep[e.g.,][]{Achitouv2017b,Paillas2021}.

For the mapping from the observed separations $s_\perp$ and $s_\parallel$ to $r$ and $\mu$, we use~\Cref{s(r)} together with~\Cref{AP} for the AP effect. This yields the following transformation between coordinates in real and redshift space,
\begin{equation}
        r_\perp = q_\perp s_\perp\;,\qquad r_\parallel = q_\parallel s_\parallel\brackets{1-\frac{1}{3}\frac{f}{b}\mathcal{M}\,\xibar(r)}^{-1}\!\;,\label{s(r)2}
\end{equation}
which can be solved via iteration to determine $r=\paren{r_\perp^2+r_\parallel^2}^{1/2}$ and $\mu=r_\parallel/r$, starting from an initial value of $r=s$~\citep{Hamaus2020}. In practice, we express all separations in units of the observable effective radius $R$ of each void in redshift space, but noting that the AP effect yields $R^*=q_\perp^{2/3}q_\parallel^{1/3}R$ in the true cosmology~\citep{Hamaus2020,Correa2021}. When expressing~\Cref{s(r)2} in units of $R^*$, only ratios of $q_\perp$ and $q_\parallel$ appear, which defines the AP parameter $\varepsilon=q_\perp/q_\parallel$. The latter is particularly well constrained via the AP test from standard spheres~\citep{Lavaux2012,Hamaus2015}.

\begin{figure*}[t]
        \centering
        \resizebox{\hsize}{!}{
                \includegraphics[trim=0 30 0 0, clip]{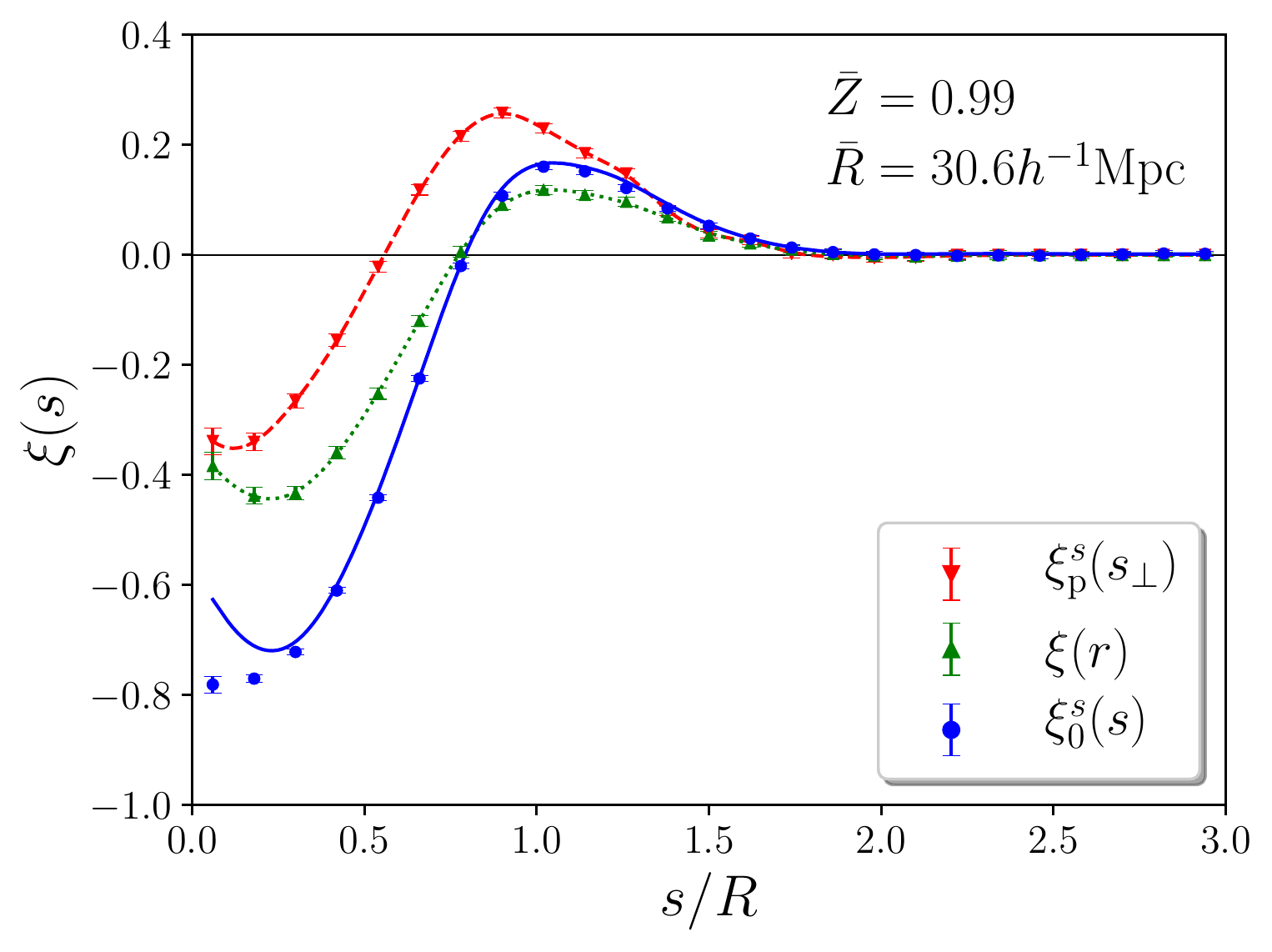}
                \includegraphics[trim=30 30 0 0, clip]{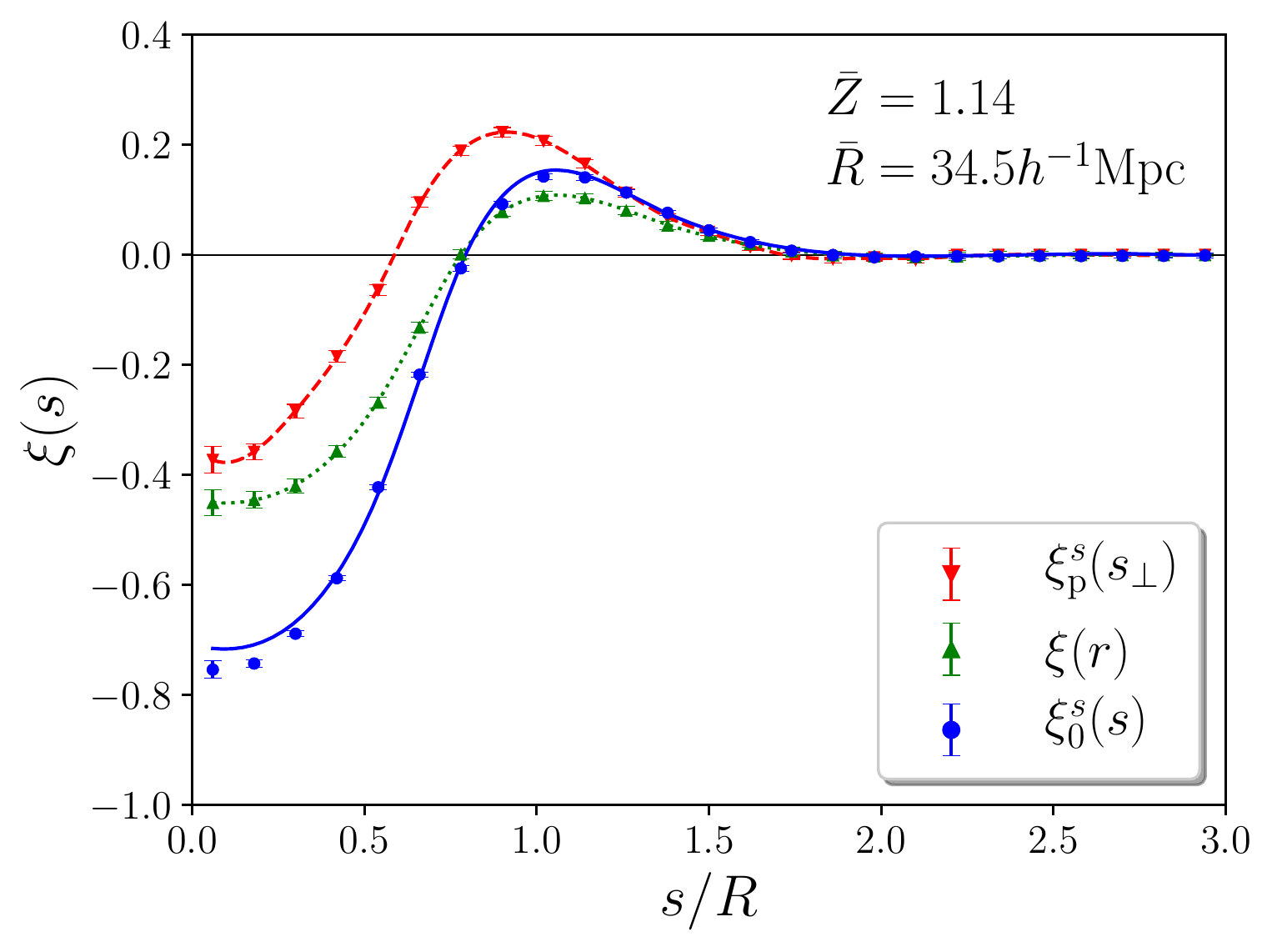}}
        \resizebox{\hsize}{!}{
                \includegraphics[trim=0 10 0 0, clip]{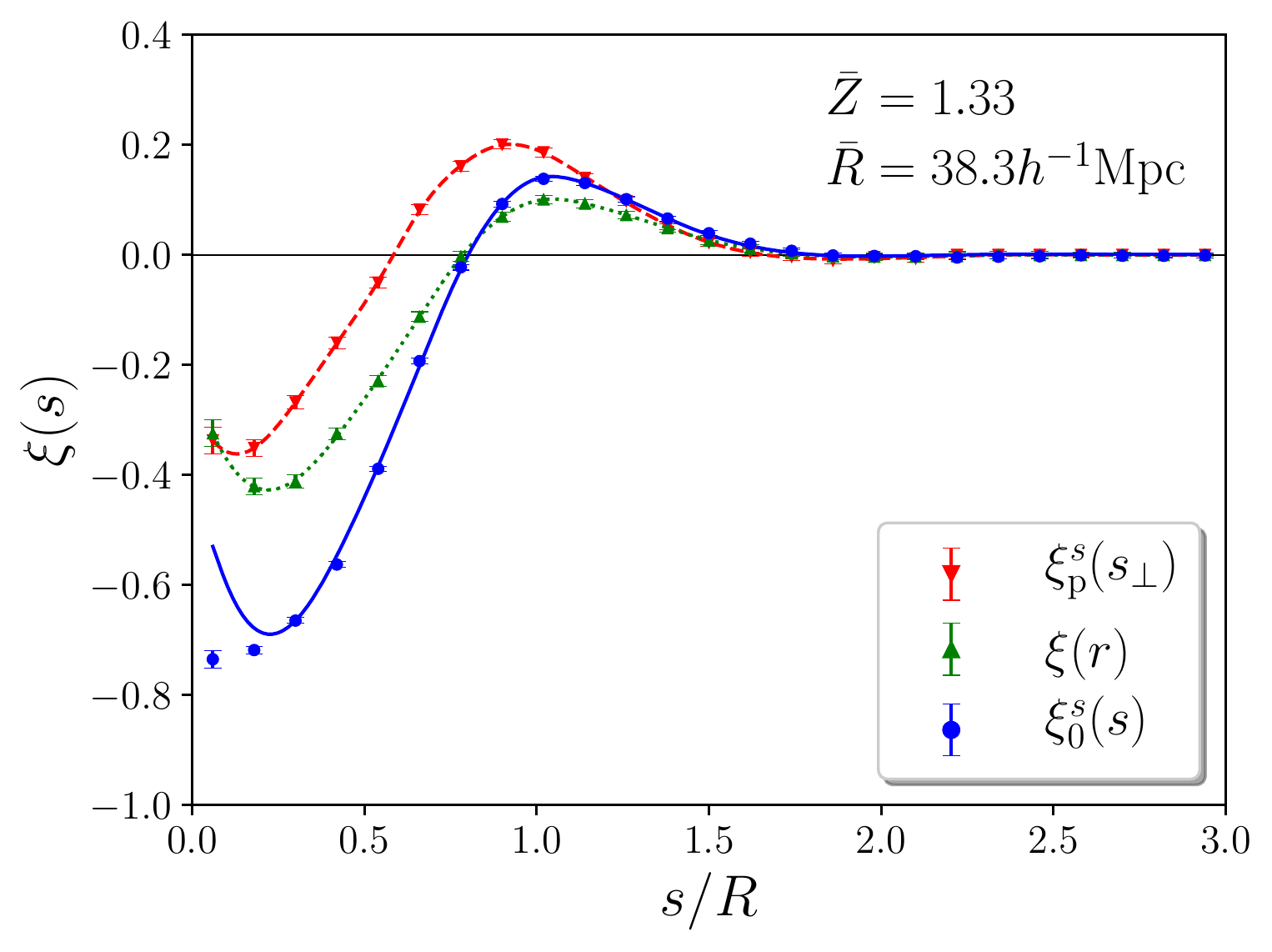}
                \includegraphics[trim=30 10 0 0, clip]{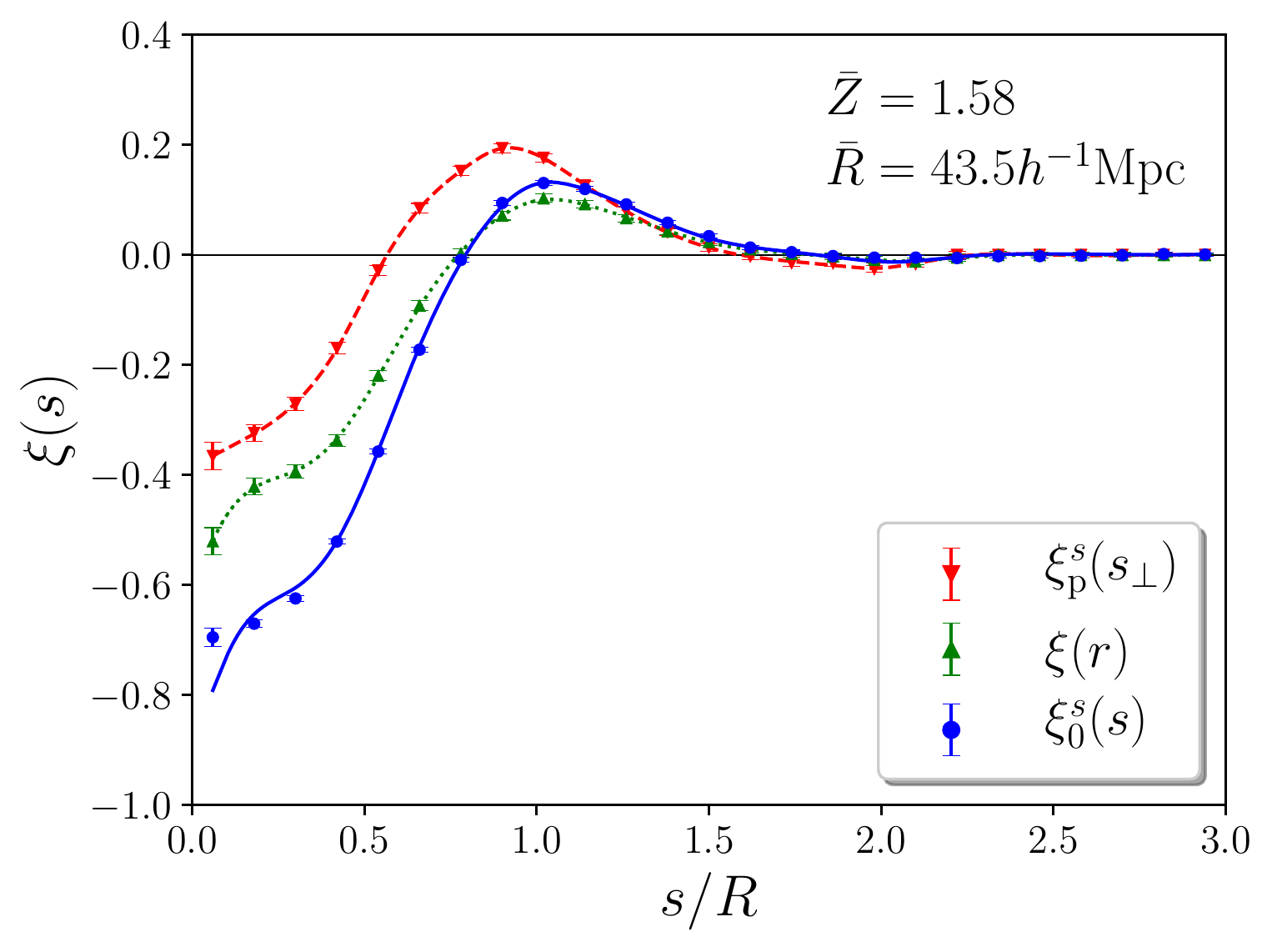}}
        \caption{Projected void-galaxy cross-correlation function $\xi^s_\mathrm{p}(s_\perp)$ in redshift space (red wedges, interpolated with dashed line) and its real-space counterpart $\xi(r)$ in 3D after deprojection (green triangles interpolated with dotted line). The redshift-space monopole $\xi^s_0(s)$ (blue dots) and its best-fit model based on \Cref{xi^s_lin3,s(r)2} is shown for comparison (solid line). Adjacent bins in redshift increase from the upper left to the lower right, with mean void redshift, $\bar{Z}$, and effective radius, $\bar{R}$, as indicated in each panel.}
        \label{fig:xip}
\end{figure*}

Finally, given the estimated data vector from~\Cref{LS_estimator}, its covariance from~\Cref{covariance}, and the model from~\Cref{xi^s_lin3,s(r)2}, we can construct a Gaussian likelihood $L(\hat{\xi}^s|\vec{\Theta})$ of the data $\hat{\xi}^s$ given the model parameter vector $\vec{\Theta}=(f/b,\varepsilon,\mathcal{M},\mathcal{Q})$ as:
\begin{equation}
        \ln L(\hat{\xi}^s|\vec{\Theta}) = -\frac{1}{2}\sum\limits_{i,j}\paren{\hat{\xi}^s(\vec{s}_i)-\xi^s(\vec{s}_i|\vec{\Theta})}\,\hat{\C}_{ij}^{-1}\paren{\hat{\xi}^s(\vec{s}_j)-\xi^s(\vec{s}_j|\vec{\Theta})}\;.
        \label{likelihood}
\end{equation}
We apply the factor of \citet{Hartlap2007} to estimate the inverse covariance matrix, which amounts to a correction of at most $3\%$ in our case. Because the model $\xi^s(\vec{s}|\vec{\Theta})$ makes use of the data to calculate $\xi(r)$ via~\Cref{xi_d}, it contributes its own covariance and becomes correlated with $\hat{\xi}^s(\vec{s})$. However, by treating the amplitude of $\xi(r)$ as the free parameter $\mathcal{M}$ in our model, we marginalize over its uncertainty. A correlation between model and data can only act to reduce the total covariance in our likelihood, so the resulting parameter errors can be regarded as upper limits in this respect. Due to their finite size, some voids extend beyond the boundaries of a given bin in redshift, which necessarily results in some degree of correlation between the parameter measurements at adjacent redshift bins. As a boundary effect, we expect this correlation to be small and hence neglect it here.

We use the affine-invariant Markov chain Monte Carlo (MCMC) ensemble sampler \texttt{emcee}~\citep{Foreman-Mackey2019} to sample the posterior probability distribution of all model parameters. The quality of the maximum-likelihood model (best fit) can be assessed via evaluation of the reduced $\chi^2$ statistic:
\begin{equation}
        \chi^2_\mathrm{red} = -\frac{2}{N_\mathrm{dof}}\ln L(\hat{\xi}^s|\vec{\Theta}) \;,
        \label{chi2}
\end{equation}
with $N_\mathrm{dof}=N_\mathrm{bin}-N_\mathrm{par}$ degrees of freedom, where $N_\mathrm{bin}$ is the number of bins for the data and $N_\mathrm{par}$ the number of model parameters. We use $18$ bins per dimension for the 2D void-galaxy cross-correlation function, which yields $N_\mathrm{bin}=324$. This number is high enough to accurately sample the scale dependence of $\hat{\xi}^s(s_\perp,s_\parallel)$, yet significantly smaller than the available number of voids per redshift bin, ensuring sufficient statistics for the estimation of its covariance matrix. With $N_\mathrm{par}=4$, this implies $N_\mathrm{dof}=320$.

\subsection{Deprojection and fit}\label{subsec:fit}
In order to evaluate our model from~\Cref{xi^s_lin3}, we need to obtain the real-space correlation function $\xi(r)$, which can be calculated via an inverse Abel transform of the projected redshift-space correlation function $\xi^s_\mathrm{p}(s_\perp)$, following~\Cref{xi_d}. We estimate the function $\xi^s_\mathrm{p}(s_\perp)$ directly via line-of-sight integration of $\hat{\xi}^s(s_\perp,s_\parallel)$, as in the first equality of~\Cref{xi_p}, and use a cubic spline to interpolate both $\xi^s_\mathrm{p}(s_\perp)$ and $\xi(r)$. The results are presented in~\Cref{fig:xip} for each of our four redshift bins. Thanks to the large number of voids in each bin, the statistical noise is low and enables a smooth deprojection of the data. Some residual noise can be noted at the innermost bins, namely, for small separations from the void center, which causes the deprojection and the subsequent spline interpolation to be less accurate~\citep{Pisani2014,Hamaus2020}. For this reason, we omit the first radial bin of the data in our model fits below, but we checked that even discarding the central three bins of $s$, $s_\perp$, and $s_\parallel$ in our analysis yields consistent results. From the full \Euclid footprint of about three times the size of an octant, the residual statistical noise of this procedure will be reduced further, so our mock analysis can be considered conservative in this regard.

We also plot the monopole of the redshift-space correlation function, which nicely follows the shape of the deprojected $\xi(r)$, as expected from~\Cref{multipoles}. Moreover, our model from~\Cref{xi^s_lin3,s(r)2} provides a very accurate fit to this monopole everywhere apart from its innermost bins, implying that any residual errors in the model remain negligibly small in that regime. We notice an increasing amplitude of all correlation functions towards lower redshift, partly reflecting the growth of overdensities along the void walls, while their interior is gradually evacuated. The increase in mean effective radius with redshift is not of dynamical origin, it is a consequence of the declining galaxy density $n_\mathrm{g}(z)$, see~\Cref{fig:nz}. A higher space density of tracers enables the identification of smaller voids.

\begin{figure*}[p]
        \centering
        \resizebox{0.87\hsize}{!}{
                \includegraphics[trim=0 32 0 6, clip]{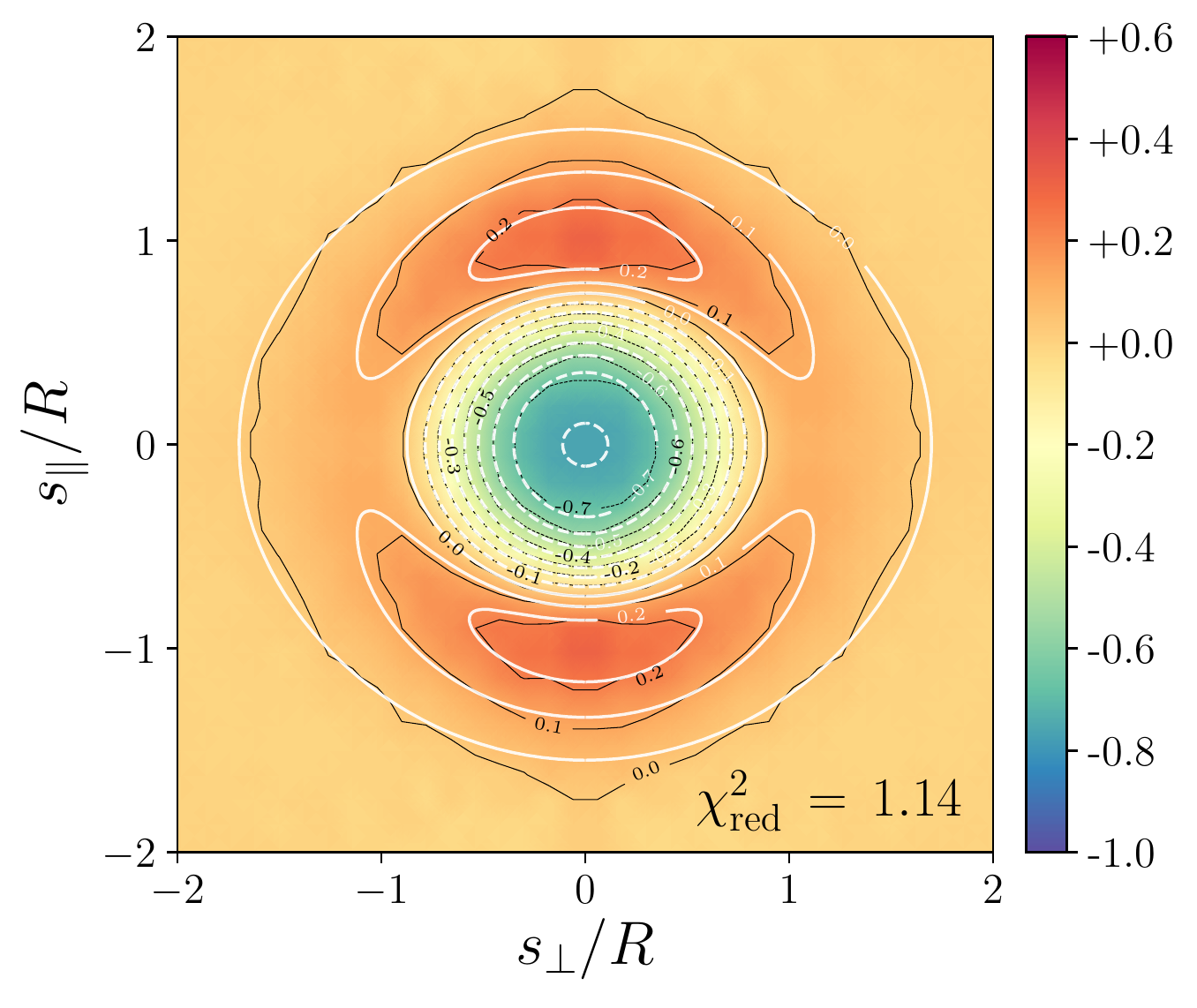}
                \includegraphics[trim=0 32 0 6, clip]{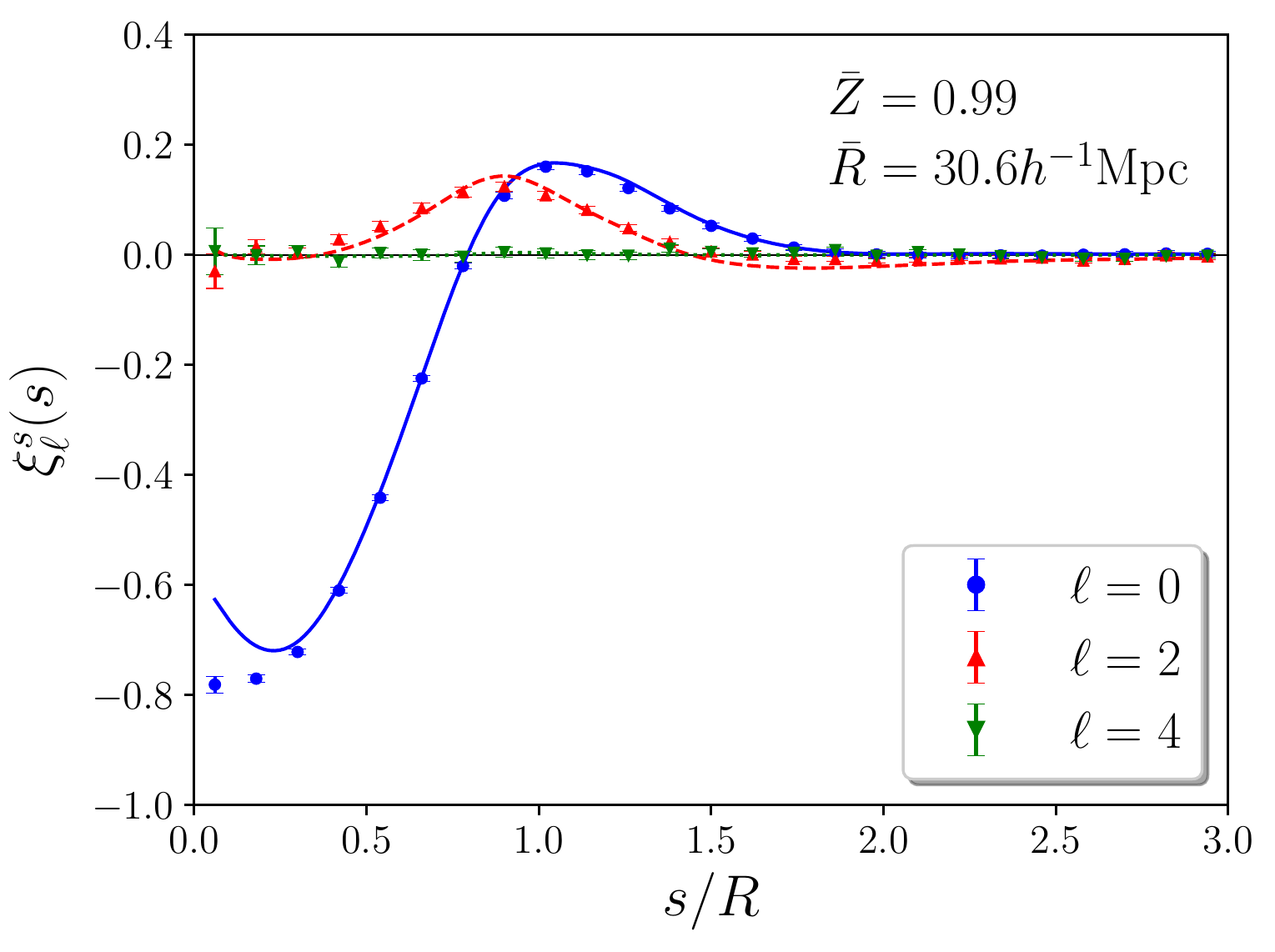}}
        \resizebox{0.87\hsize}{!}{
                \includegraphics[trim=0 32 0 6, clip]{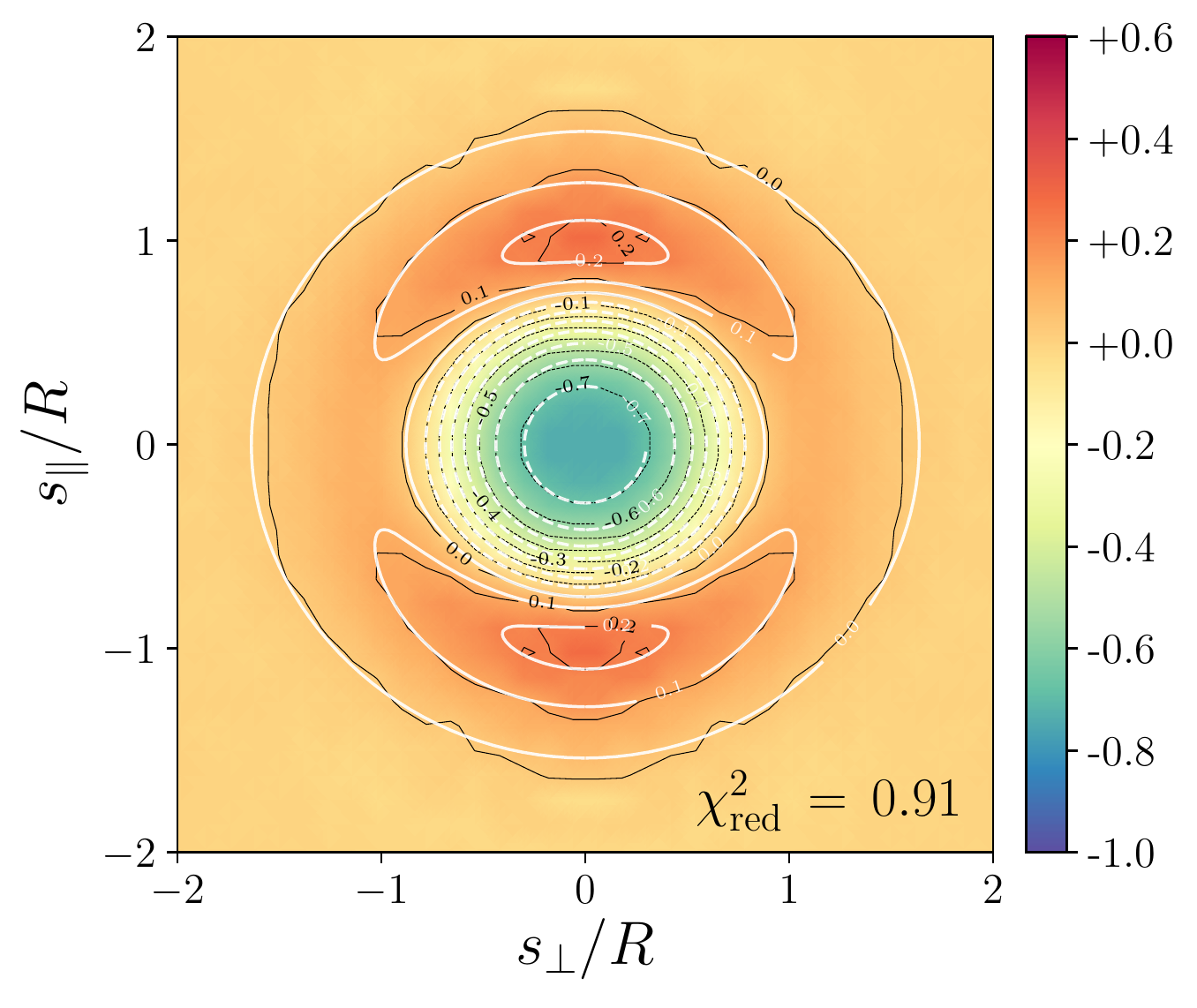}
                \includegraphics[trim=0 32 0 6, clip]{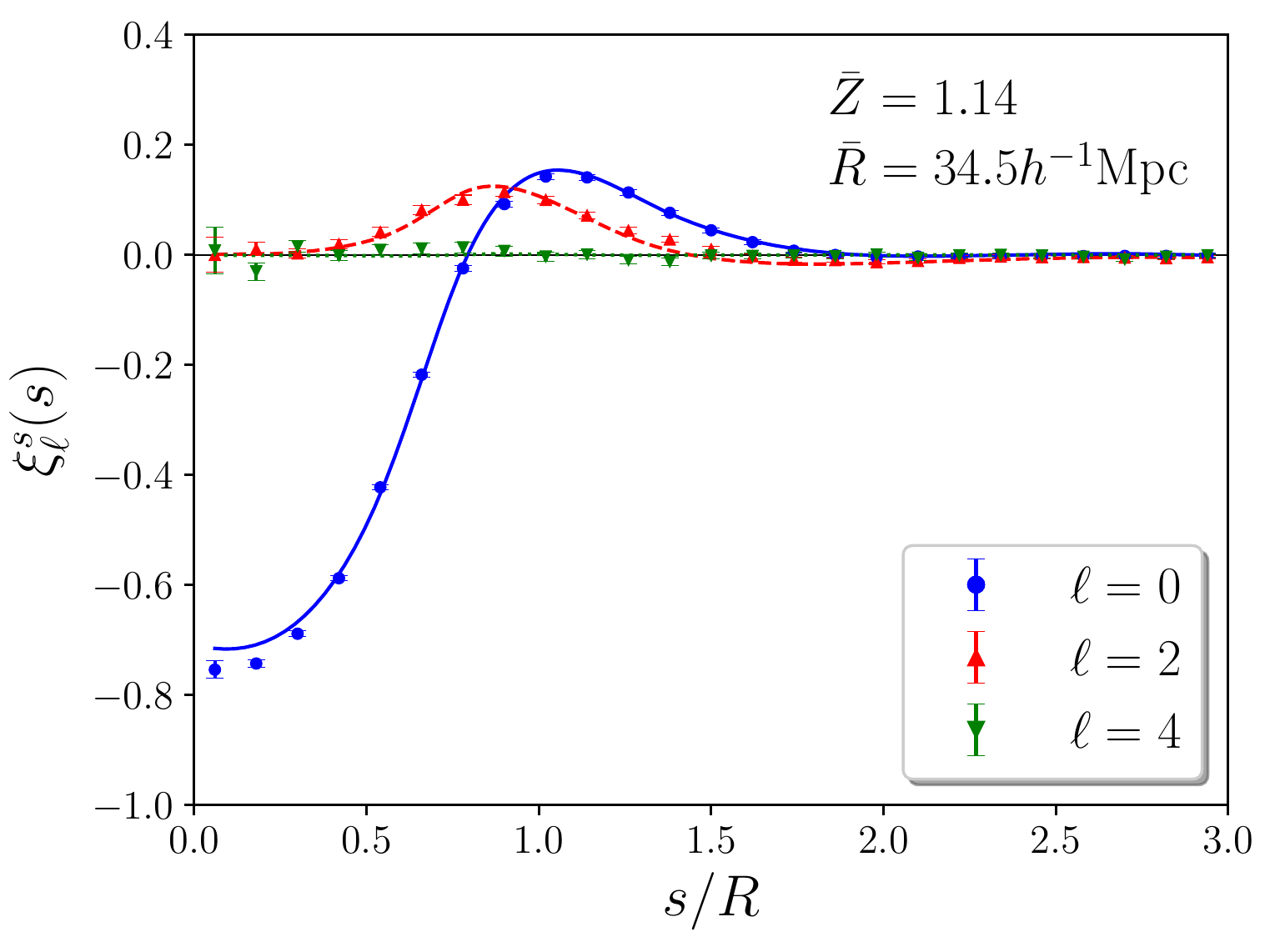}}
        \resizebox{0.87\hsize}{!}{
                \includegraphics[trim=0 32 0 6, clip]{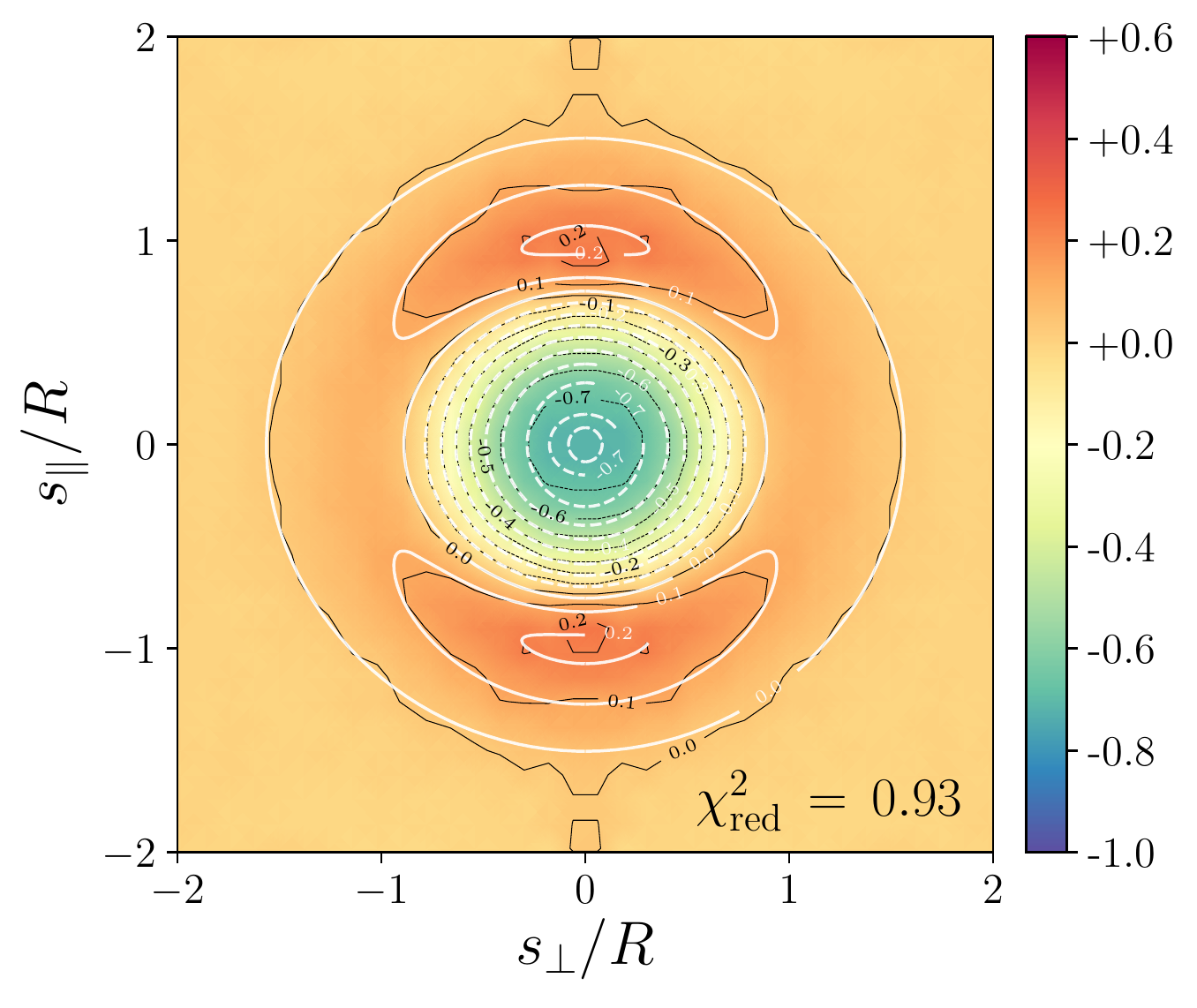}
                \includegraphics[trim=0 32 0 6, clip]{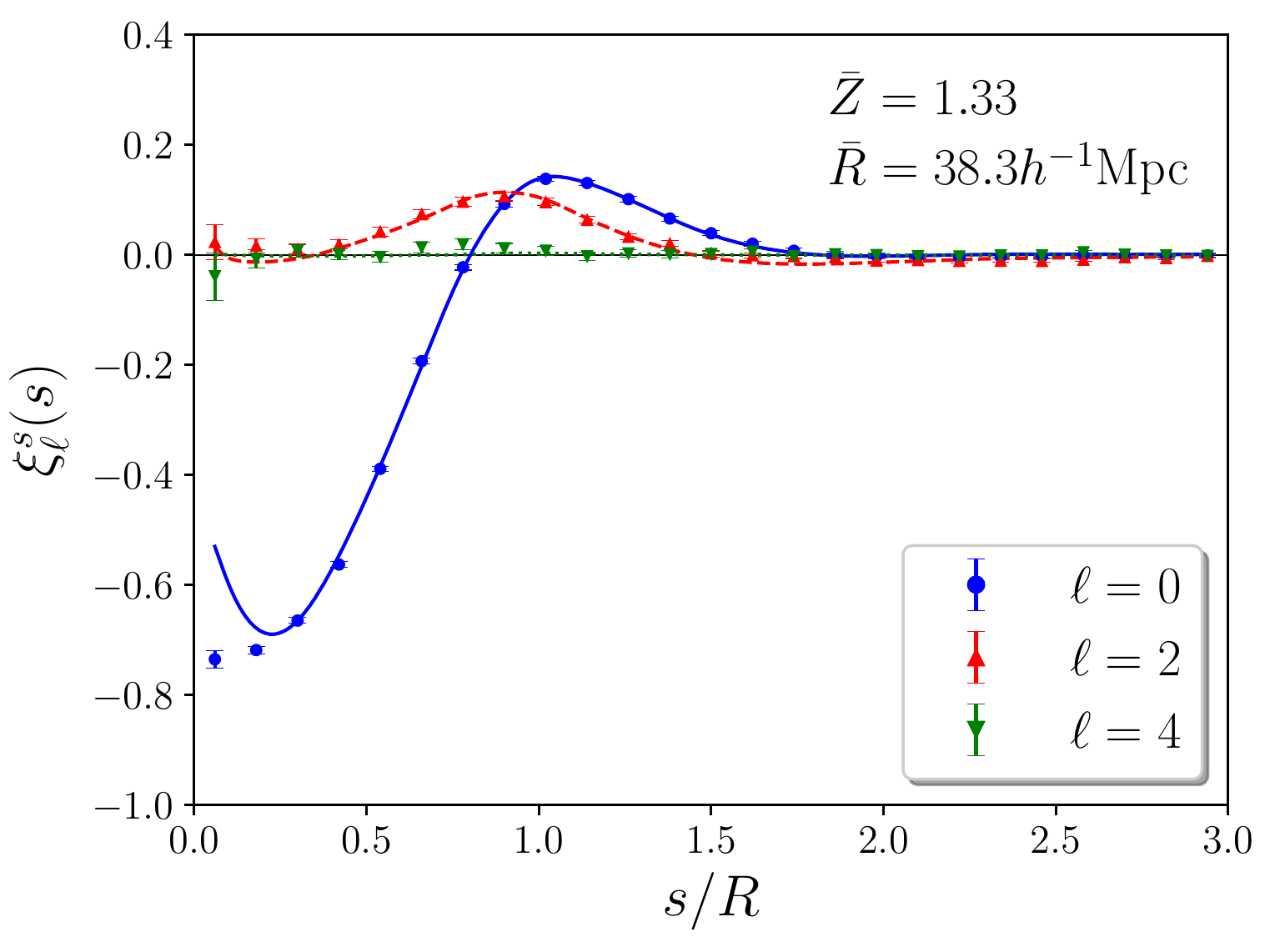}}
        \resizebox{0.87\hsize}{!}{
                \includegraphics[trim=0 10 0 6, clip]{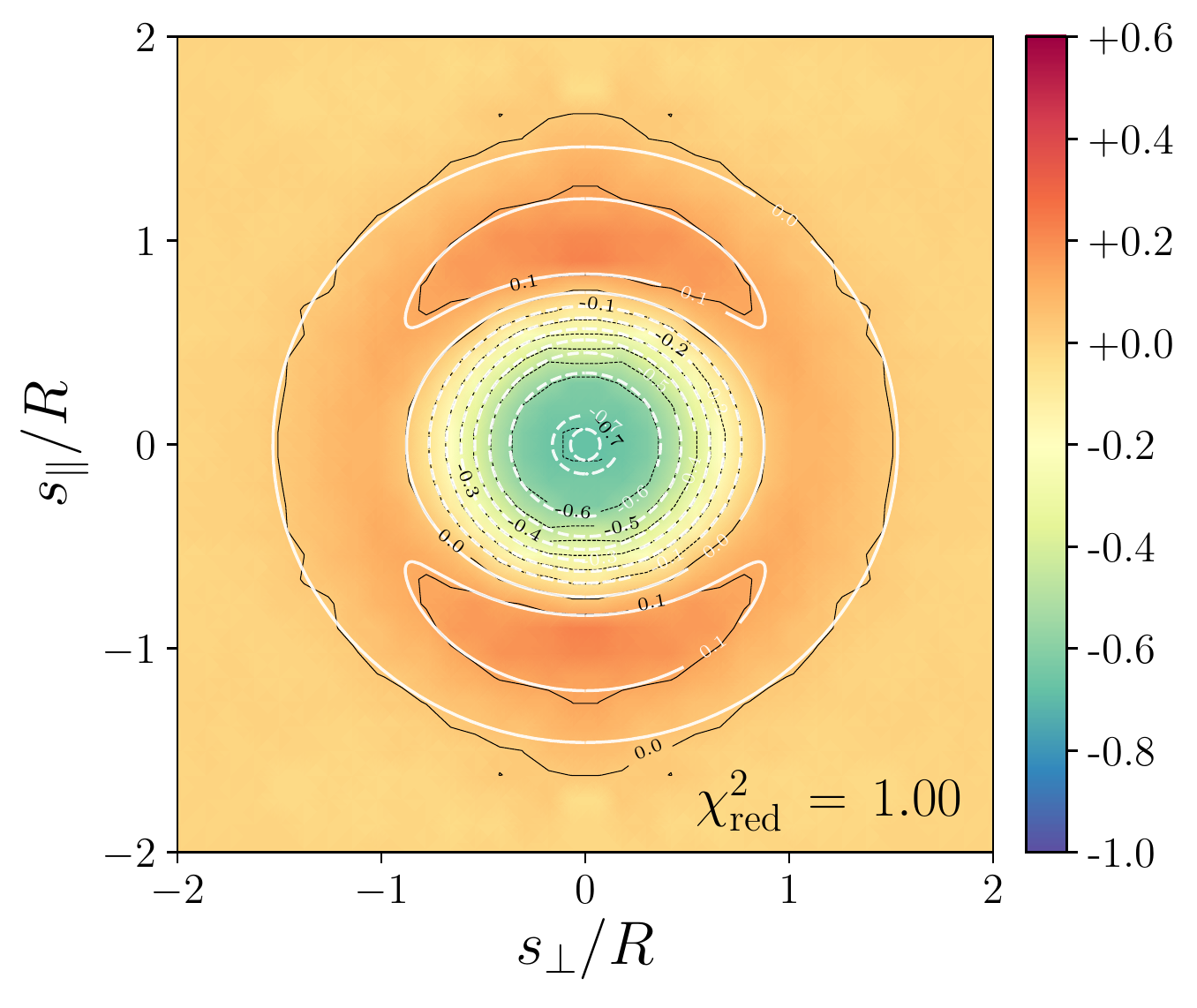}
                \includegraphics[trim=0 10 0 6, clip]{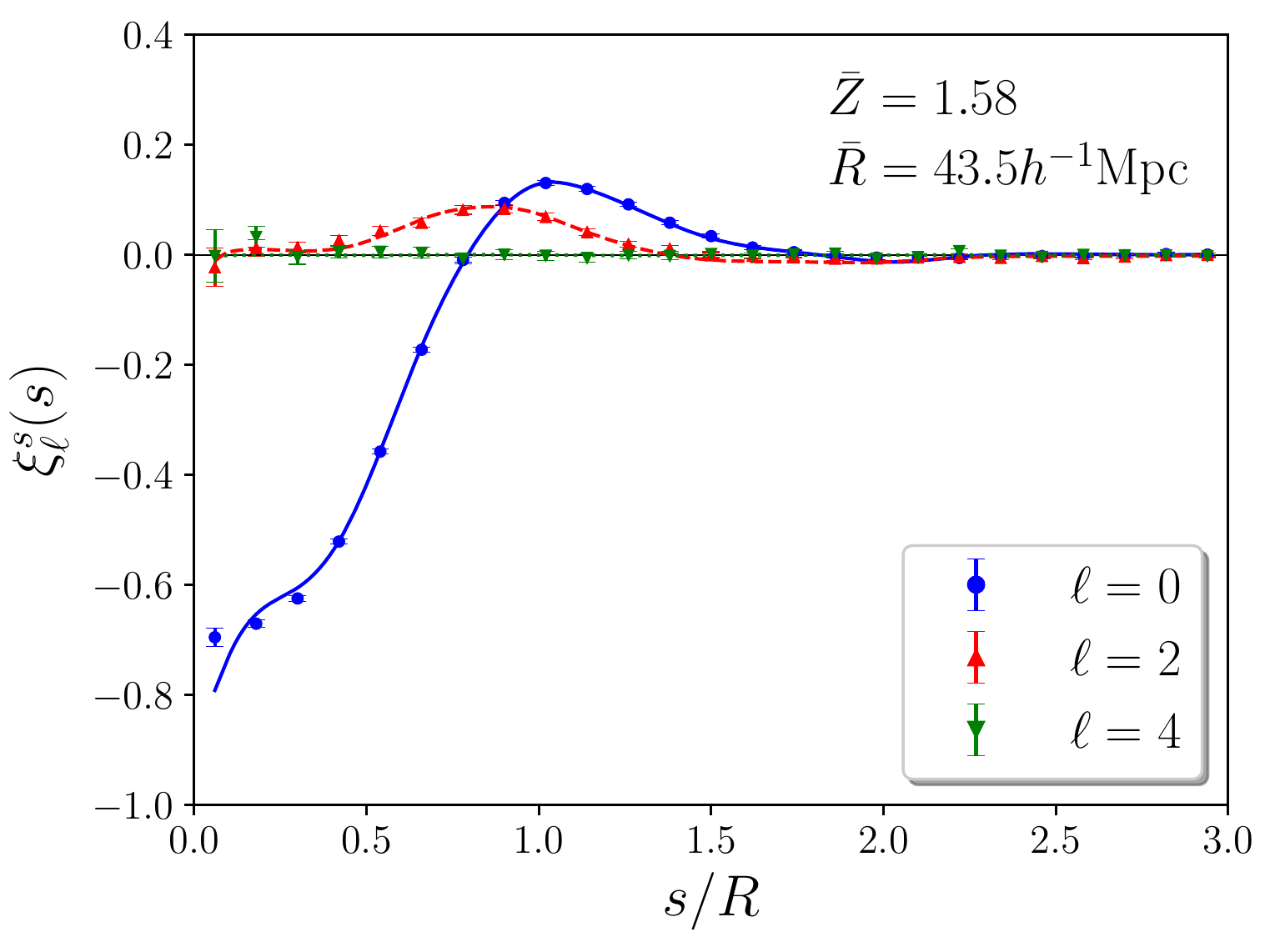}}
        \caption{Stacked void-galaxy cross-correlation function in redshift space. \textit{Left}: $\xi^s(s_\perp,s_\parallel)$ in 2D (color scale with black contours) and its best-fit model from~\Cref{xi^s_lin3,s(r)2} (white contours). \textit{Right}: Monopole (blue dots), quadrupole (red triangles) and hexadecapole (green wedges) of $\xi^s(s_\perp,s_\parallel)$ and best-fit model (solid, dashed, dotted lines). The mean void redshift, $\bar{Z}$, and effective radius, $\bar{R}$, of each redshift bin are indicated.}
        \label{fig:xi2d}
\end{figure*}

\begin{figure*}[p]
        \centering
        \resizebox{0.9\hsize}{!}{
                \includegraphics[trim=0 0 0 0, clip]{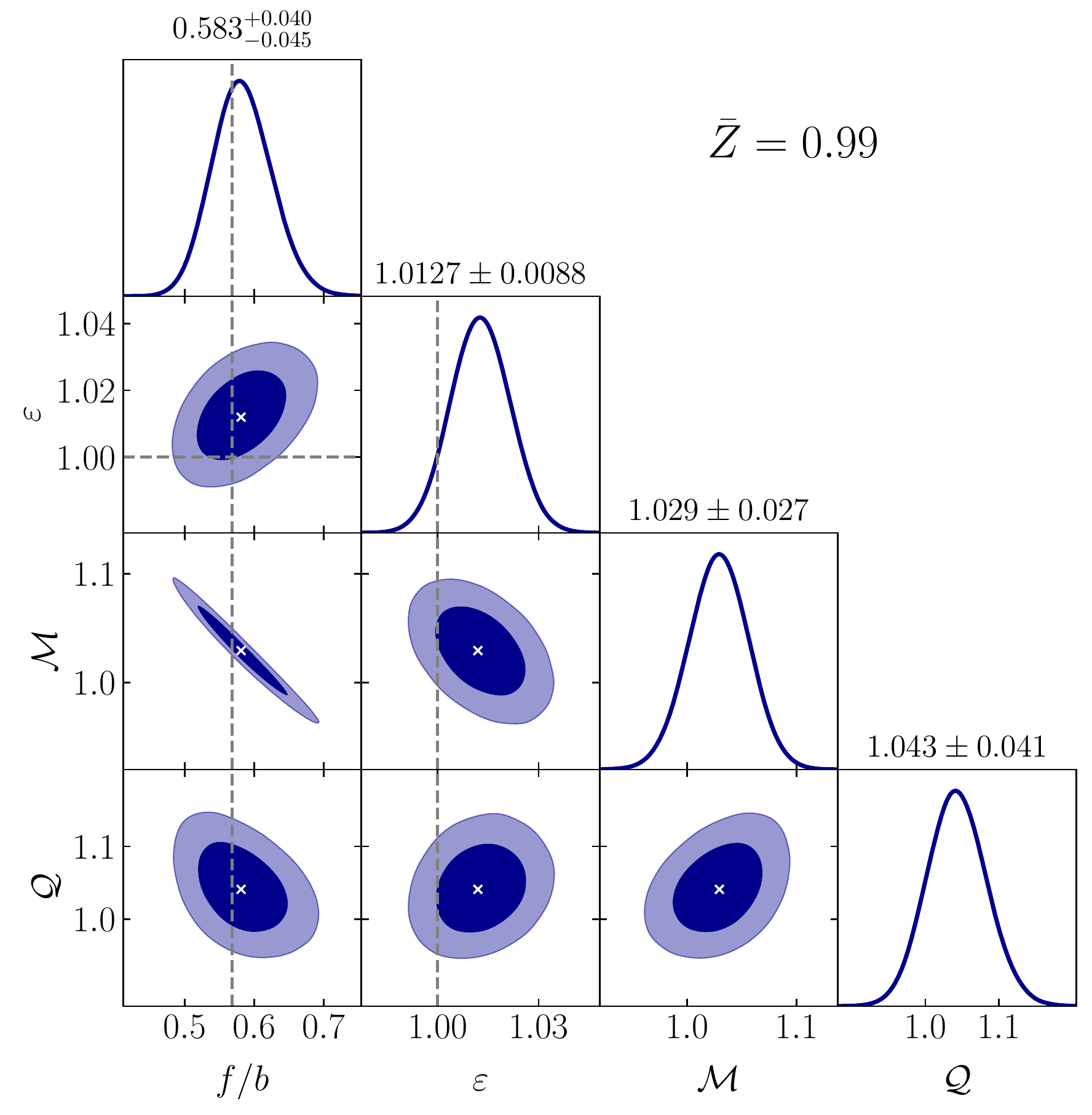}
                \includegraphics[trim=0 0 0 0, clip]{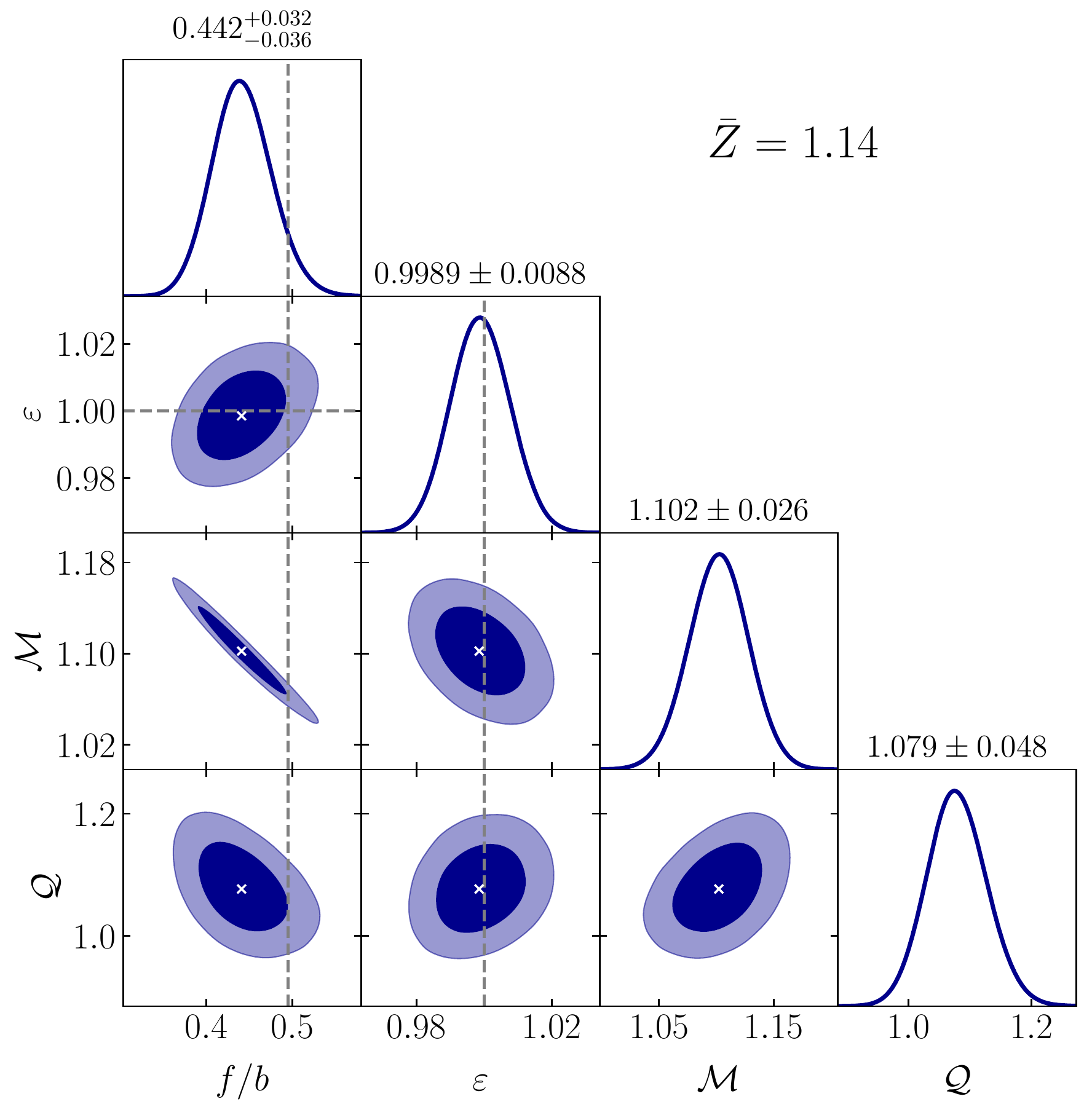}}
        \resizebox{0.9\hsize}{!}{
                \includegraphics[trim=8 0 0 0, clip]{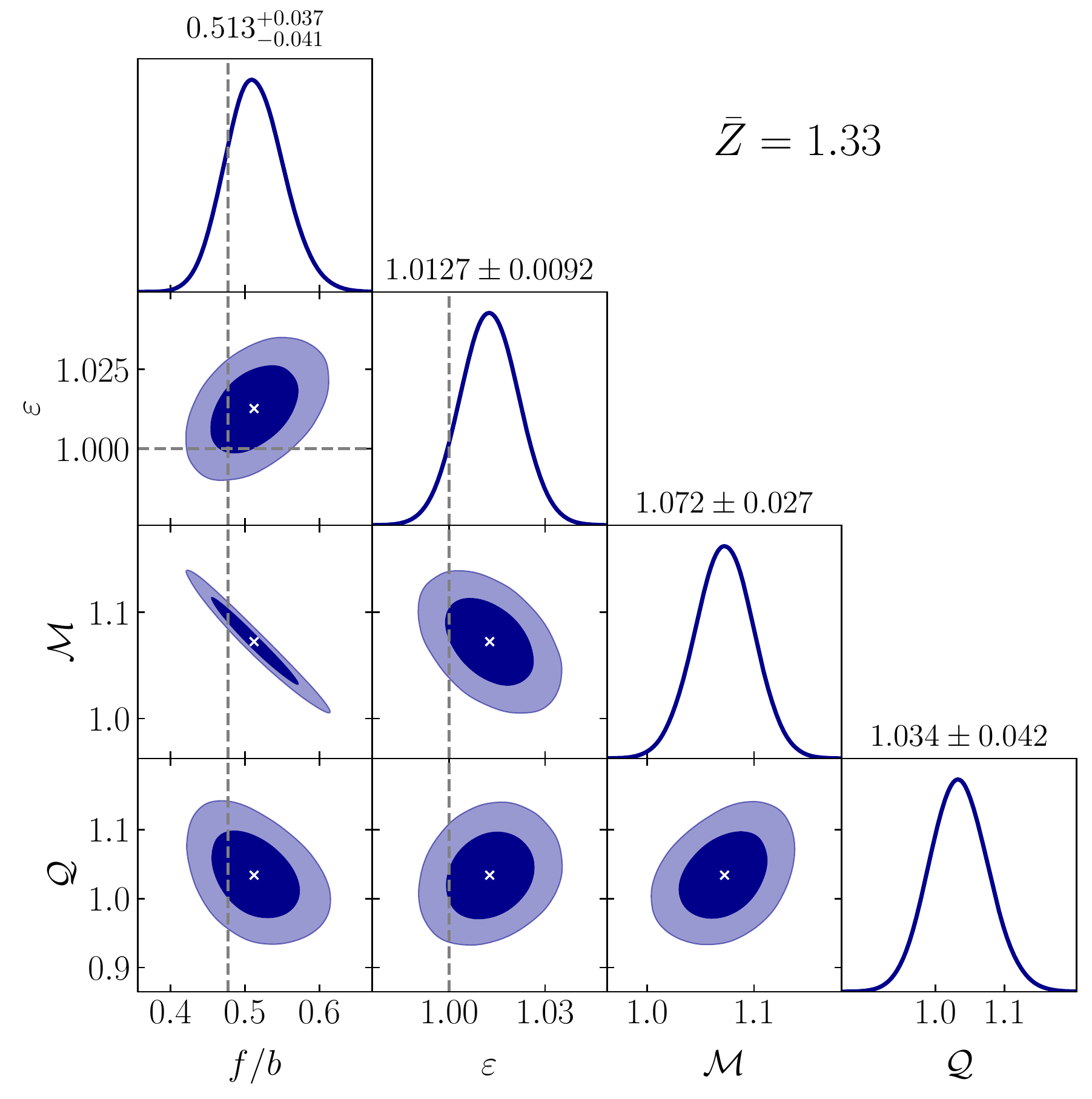}
                \includegraphics[trim=0 0 0 0, clip]{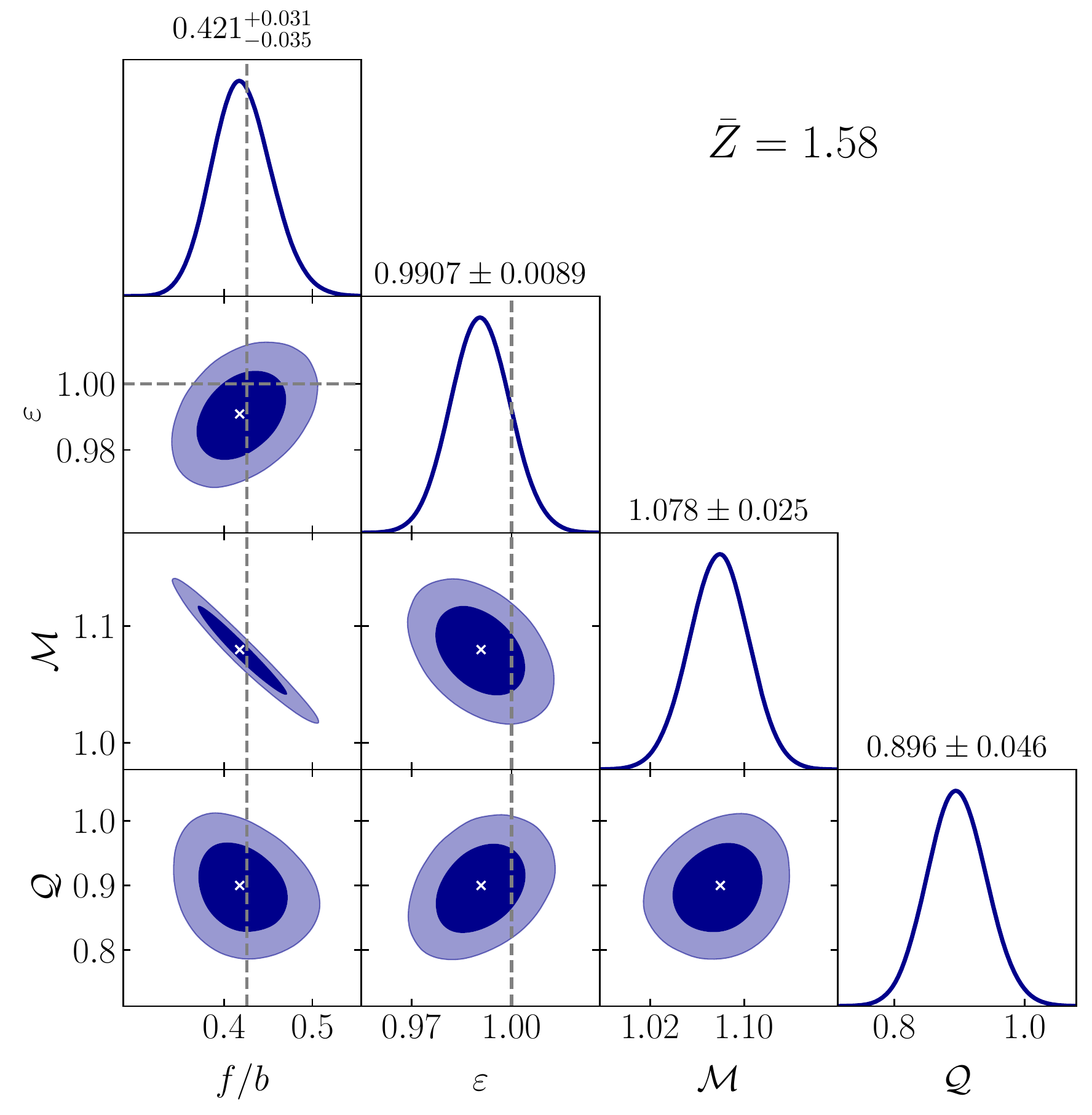}}
        \caption{Posterior probability distribution of the model parameters that enter in~\Cref{xi^s_lin3,s(r)2}, obtained via MCMC from the data shown in the left of~\Cref{fig:xi2d}. Dark and light shaded areas represent $68\%$ and $95\%$ confidence regions with a cross marking the best fit, dashed lines indicate fiducial values of the RSD and AP parameters. The top of each column states the mean and standard deviation of the 1D marginal distributions. Adjacent bins in void redshift with mean value $\bar{Z}$ increase from the upper left to the lower right, as indicated.}
        \label{fig:triangle}
        \vspace{15pt}
        \resizebox{0.95\hsize}{!}{
                \includegraphics[trim=10 1 0 0, clip]{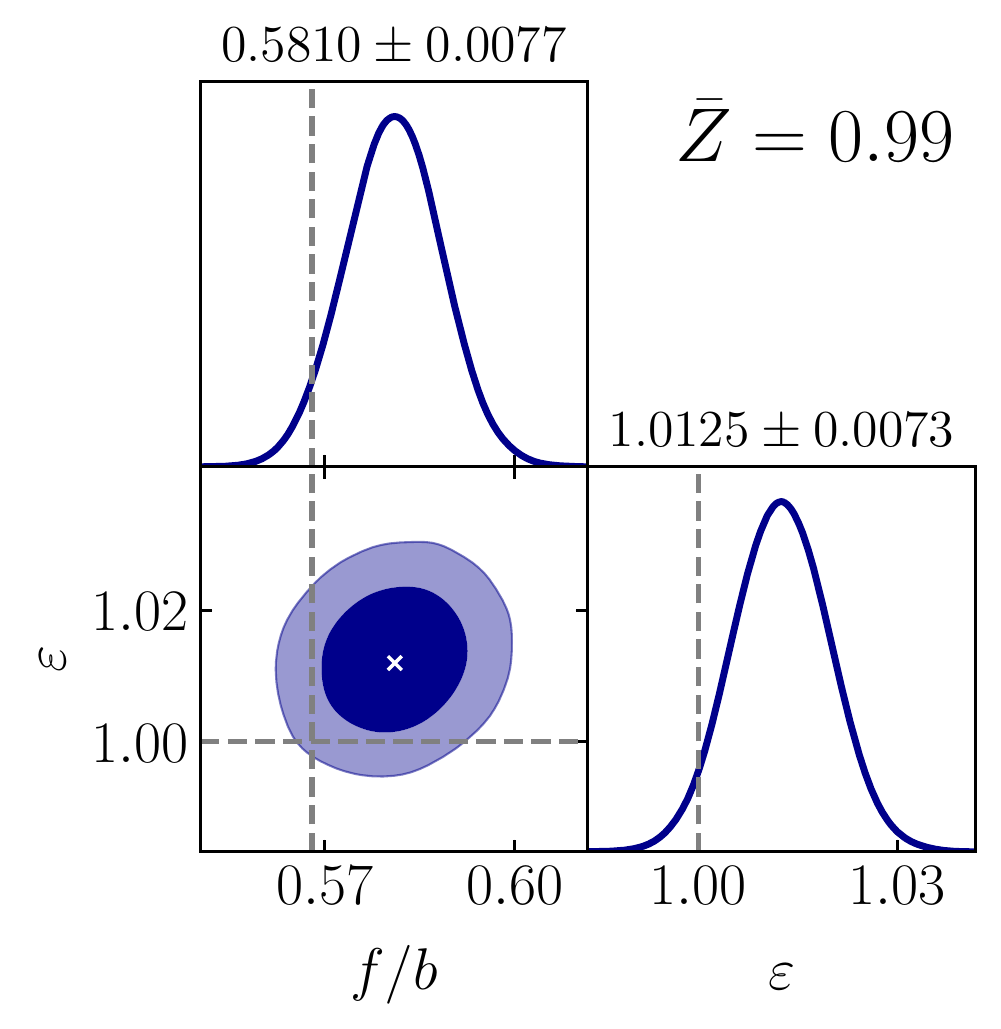}
                \includegraphics[trim=10 1 0 0, clip]{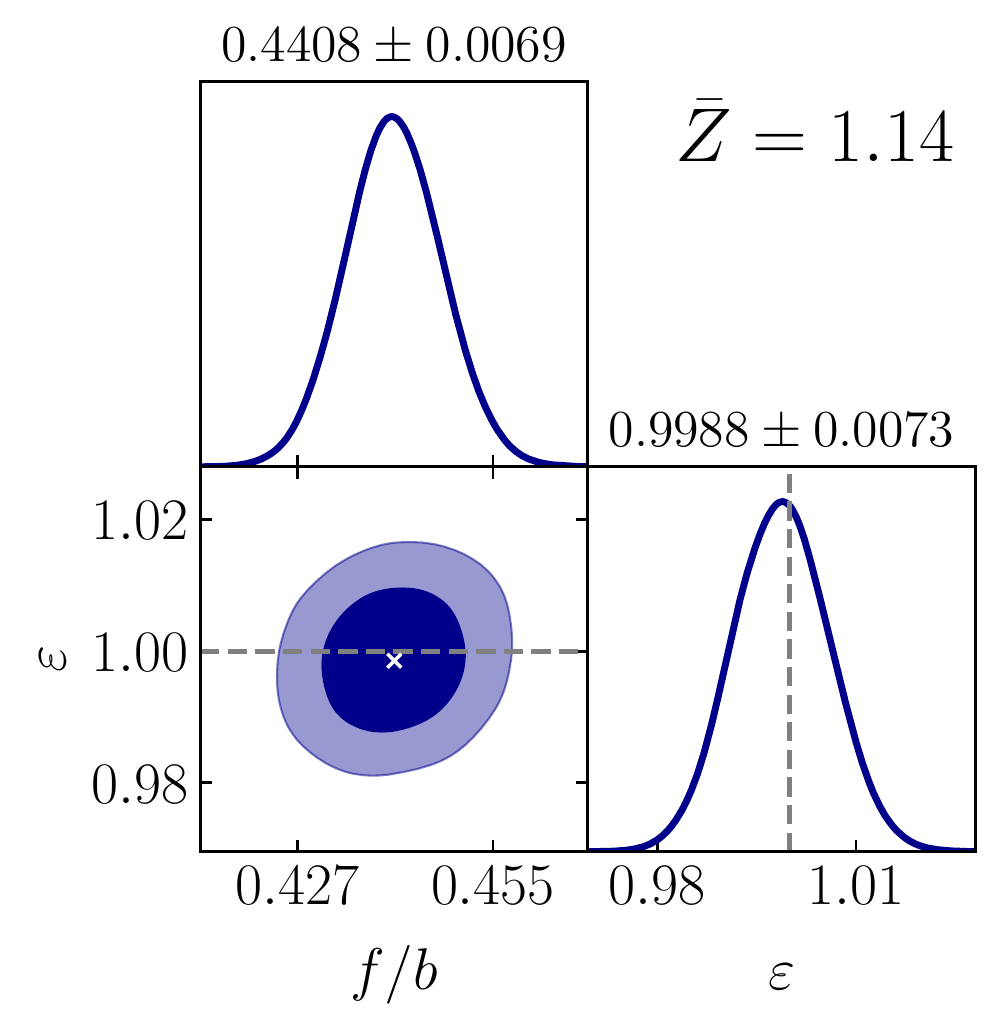}
                \includegraphics[trim=10 1 0 0, clip]{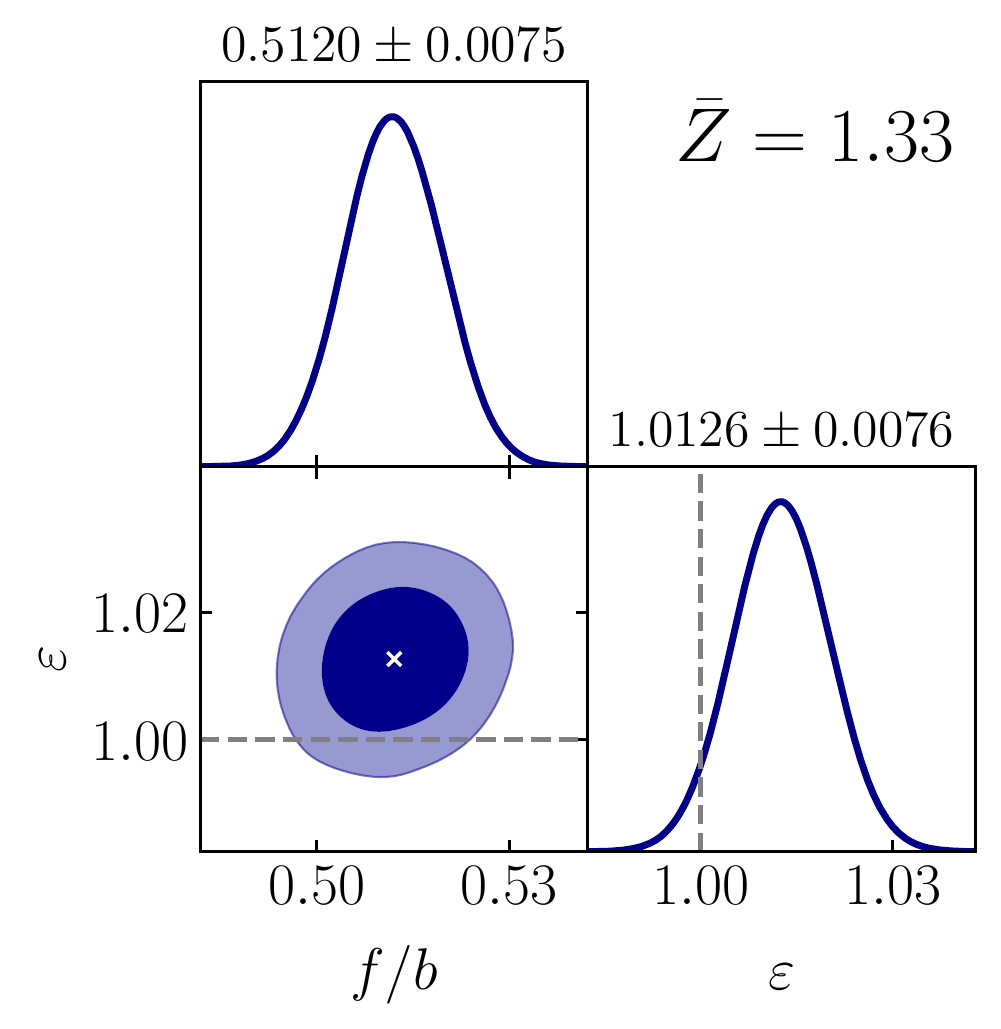}
                \includegraphics[trim=10 1 0 0, clip]{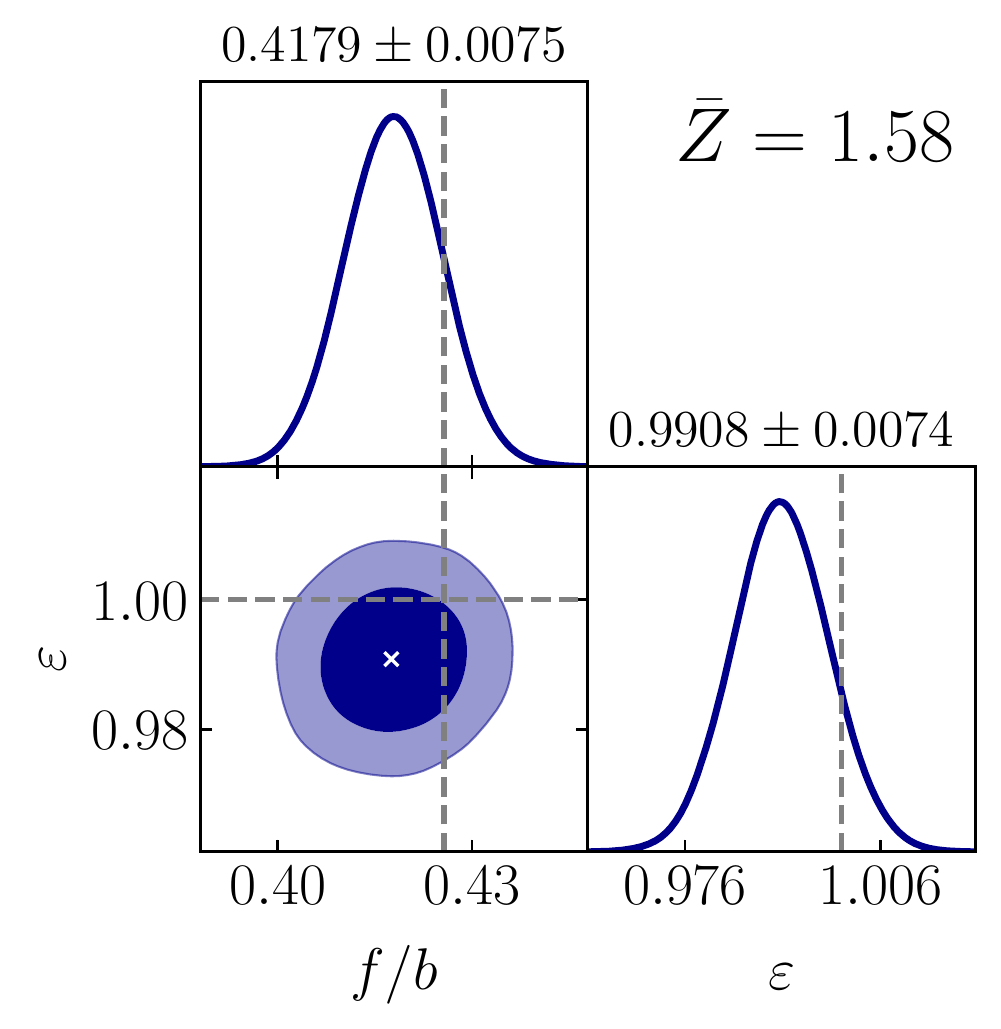}}
        \caption{Posterior probability distribution of the model parameters that enter in~\Cref{xi^s_lin3,s(r)2}. Details are the same as in~\Cref{fig:triangle}, but with a fixing (calibration) of the nuisance parameters $\mathcal{M}$ and $\mathcal{Q}$ to their best-fit values. Redshift increases from left to right, as indicated.}
        \label{fig:triangle_cal}
\end{figure*}

We minimize the log-likelihood of~\Cref{likelihood} by varying the model parameters to find the best-fit model to the mock data. As a data vector, we used the 2D void-galaxy cross-correlation $\hat{\xi}^s(s_\perp,s_\parallel)$, which contains the complete information on dynamic and geometric distortions from all multipole orders. However, we checked that our pipeline yields consistent constraints when only considering the three lowest even multipoles, $\hat{\xi}^s_\ell(s)$, of order $\ell=0,2,4$ as our data vector. The results are shown in~\Cref{fig:xi2d} for our four consecutive redshift bins. In each case, we find extraordinary agreement between the model and the data, which is further quantified by the reduced $\chi^2$ being so close to unity in all bins. Again, it is possible to perceive the slight deepening of voids over time and the agglomeration of matter in their surroundings. The multipoles shown in the right column of~\Cref{fig:xi2d} complement this view, with both monopole and quadrupole enhancing their amplitudes during void evolution, and they exhibit an excellent agreement with the model. The quadrupole vanishes towards the central void region and the model mismatch in the first few bins of the monopole has negligible impact here, as most of the anisotropic information originates from larger scales. In addition, the hexadecapole remains consistent with zero at all times, in accordance with~\Cref{multipoles}.

\begin{table*}[ht]
        \caption{RSD and AP parameter constraints}\label{tab:constraints}
        \centering
        \centerline{
                \begin{tabular}{lccccccccc}
                        \toprule
                         Data & & $Z_\mathrm{min}$ & $Z_\mathrm{max}$ & $\bar{Z}$ & $b$ & $f/b$  & $f\sigma_8$ & $\varepsilon$ & $\DA H/c$ \\
                        \midrule
                         \Euclid voids & & $0.91$ & $1.06$ & $0.99$ & $1.54$ & $0.5827\pm0.0427$ & $0.4544\pm0.0333$ & $1.0127\pm0.0088$ & $1.3627\pm0.0119$ \\[3pt]
                         (independent) & & $1.06$ & $1.23$ & $1.14$ & $1.81$ & $0.4416\pm0.0346$ & $0.3784\pm0.0296$ & $0.9989\pm0.0088$ & $1.6391\pm0.0144$ \\[3pt]
                                       & & $1.23$ & $1.44$ & $1.33$ & $1.92$ & $0.5132\pm0.0394$ & $0.4320\pm0.0331$ & $1.0127\pm0.0092$ & $2.0517\pm0.0186$ \\[3pt]
                                       & & $1.44$ & $1.76$ & $1.58$ & $2.20$ & $0.4205\pm0.0337$ & $0.3689\pm0.0296$ & $0.9907\pm0.0089$ & $2.5543\pm0.0230$ \\
                        \midrule
                         \Euclid voids & & $0.91$ & $1.06$ & $0.99$ & $1.54$ & $0.5810\pm0.0076$ & $0.4531\pm0.0060$ & $1.0125\pm0.0073$ & $1.3624\pm0.0098$ \\[3pt]
                         (calibrated)  & & $1.06$ & $1.23$ & $1.14$ & $1.81$ & $0.4408\pm0.0069$ & $0.3777\pm0.0059$ & $0.9988\pm0.0073$ & $1.6389\pm0.0120$ \\[3pt]
                                       & & $1.23$ & $1.44$ & $1.33$ & $1.92$ & $0.5120\pm0.0075$ & $0.4310\pm0.0064$ & $1.0127\pm0.0076$ & $2.0518\pm0.0153$ \\[3pt]
                                       & & $1.44$ & $1.76$ & $1.58$ & $2.20$ & $0.4179\pm0.0075$ & $0.3666\pm0.0066$ & $0.9908\pm0.0074$ & $2.5546\pm0.0191$ \\
                        \bottomrule
        \end{tabular}}
        \tablefoot{Forecasted constraints on RSD and AP parameters (mean values with $68\%$ confidence intervals) from \texttt{VIDE} voids in the \Euclid Flagship mock catalog (top rows). Results are given in four redshift bins with minimum, maximum, and mean void redshift $Z_\mathrm{min}$, $Z_\mathrm{max}$, $\bar{Z}$, and large-scale galaxy bias $b$. All uncertainties correspond to one octant of the sky, but the expected precision from the full \Euclid footprint is a factor of about $\!\sqrt{3}$ higher. The bottom rows show more optimistic constraints after performing a calibration of the nuisance parameters in the model.}
        \vspace{10pt}
\end{table*}

\begin{figure*}[ht]
        \centering
        \resizebox{0.95\hsize}{!}{
                \includegraphics[trim=15 5 5 5]{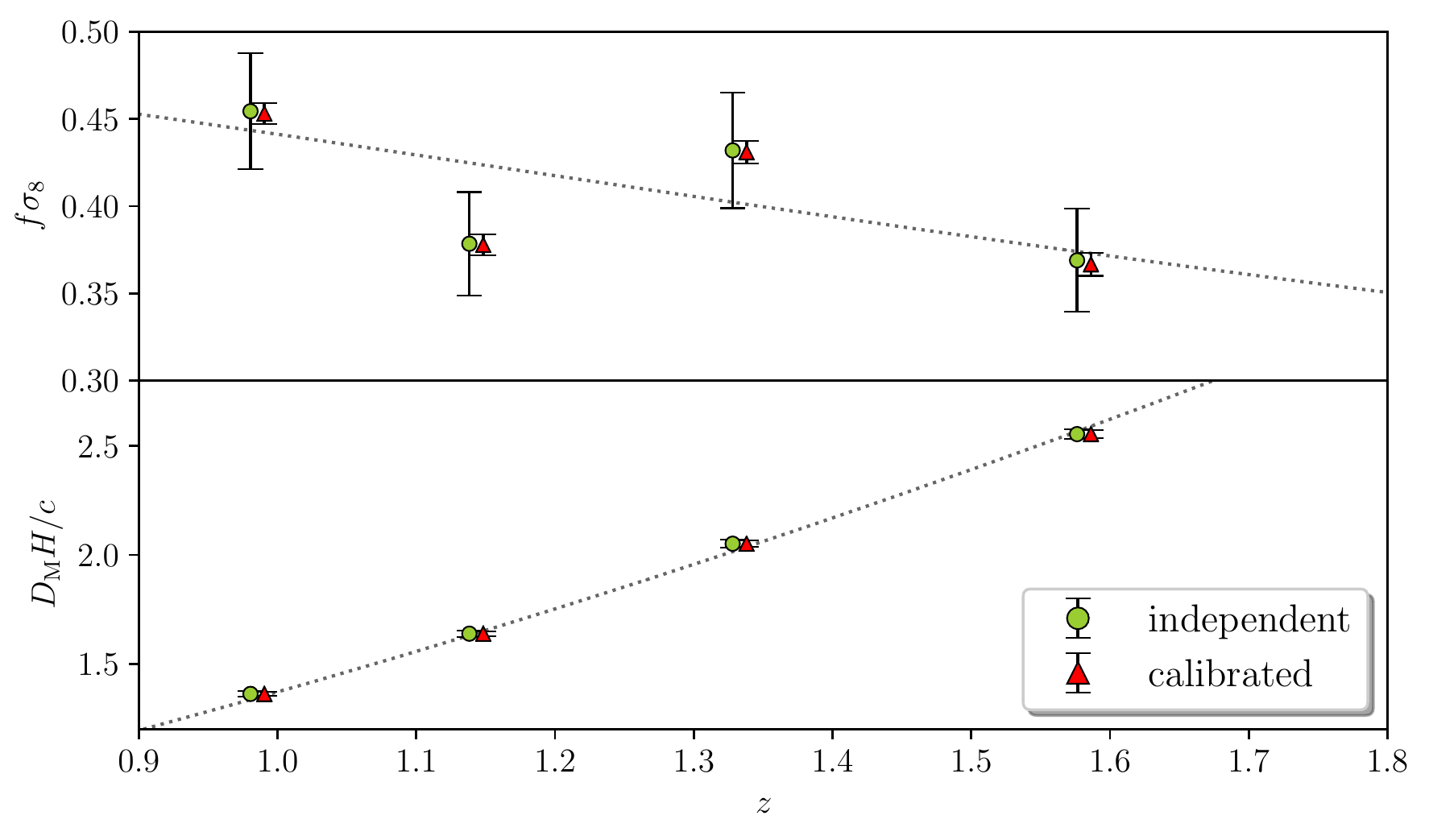}}
        \caption{Measurement of $f\sigma_8$ and $\DA H$ from \texttt{VIDE} voids in the \Euclid Flagship catalog as a function of redshift $z$. The marker style distinguishes between a fully model-independent approach (green circles) and an analysis with calibrated nuisance parameters $\mathcal{M}$ and $\mathcal{Q}$ (red triangles) using external sources, such as simulations or mocks. Dotted lines indicate the Flagship input cosmology, the markers are slightly shifted horizontally for visibility. These results are based on one octant of the sky, the expected precision from the full \Euclid footprint is a factor of about $\!\sqrt{3}$ higher.}
        \label{fig:fs8_DAH}
\end{figure*}

\subsection{Parameter constraints}\label{subsec:constraints}
Performing a full MCMC run for each redshift bin, we obtain the posteriors for our model parameters as shown in~\Cref{fig:triangle}. We observe a similar correlation structure as in the previous BOSS analysis of~\cite{Hamaus2020}. Namely, a weak correlation between $f/b$ and $\varepsilon$, and a strong anti-correlation between $f/b$ and $\mathcal{M}$. Overall, the $68\%$ confidence regions for $f/b$ and $\varepsilon$ agree well with the expected input values, shown as dashed lines. In a \lcdm\ cosmology, the linear growth rate is given by
\begin{equation}
        f(z)\simeq\brackets{\frac{\Om(1+z)^3}{H^2(z)/H_0^2}}^{\,\gamma},
        \label{growth_rate}
\end{equation}
with a growth index of $\gamma\simeq0.55$~\citep{Lahav1991,Linder2005}. We take the Flagship measurements of the large-scale linear bias $b$ from our companion paper~(\textcolor{blue}{Contarini et al. in prep.}), which uses \texttt{CAMB}~\citep{Lewis2000} to calculate the dark-matter correlation function to compare with the estimated galaxy auto-correlation function~\citep[see][for details]{Marulli2013,Marulli2018}. As we are using the correct input cosmology of Flagship to convert angles and redshifts to comoving distances following~\Cref{comoving}, we have $\varepsilon=1$ as fiducial value.

For the values of the nuisance parameters, $\mathcal{M}$ and $\mathcal{Q}$, we do not have any specific expectation, but we set their defaults to unity as well. We also find their posteriors to be distributed around values of one, although their mean can deviate more than one standard deviation from that default value. However, the distributions of the nuisance parameters are not relevant for the cosmological interpretation of the posterior, as they can be marginalized over. The relative precision on $f/b$ ranges between $7.3\%$ and $8.0\%$, while the one on $\varepsilon$ is between $0.87\%$ and $0.91\%$. This precision corresponds to a survey area of one octant of the sky, but the footprint covered by \Euclid will be roughly three times as large. Therefore, one can expect these numbers to decrease by a factor of $\!\sqrt{3}$ to yield about $4\%$ accuracy on $f/b$ and $0.5\%$ on $\varepsilon$ per redshift bin.

The attainable precision can even further be increased via a calibration strategy. \cite{Hamaus2020} have shown that this is possible when the model ingredients $\xi(r)$, $\mathcal{M}$, and $\mathcal{Q}$ are taken from external sources, instead of being constrained by the data itself, for example, from a large number of high-fidelity survey mocks. However, we emphasize that this practice introduces a prior dependence on the assumed model parameters in the mocks, so it underestimates the final uncertainty on cosmology and may yield biased results. We also note that survey mocks are typically designed to reproduce the two-point statistics of galaxies on large scales, but are not guaranteed to provide void statistics at a similar level of accuracy.

Nevertheless, for the sake of completeness we investigate the achievable precision when fixing the nuisance parameters to their best-fit values in the full analysis, while still inferring $\xi(r)$ via deprojection of the data as before. We note that this is an arbitrary choice of calibration, in practice, the values of $\mathcal{M}$ and $\mathcal{Q}$ will depend on the type of mocks considered. The resulting posteriors on $f/b$ and $\varepsilon$ are shown in~\Cref{fig:triangle_cal}. The calibrated analysis yields a relative precision of $1.3\%$ to $1.8\%$ on $f/b$ and $0.72\%$ to $0.75\%$ on $\varepsilon$. Compared to the calibration-independent analysis, this amounts to an improvement by up to a factor of about $5$ for constraints on $f/b$ and $1.2$ for $\varepsilon$. Extrapolated to the full survey area accessible to \Euclid, this corresponds to a precision of roughly $1\%$ on $f/b$ and $0.4\%$ on $\varepsilon$ per redshift bin. As expected, these calibrated constraints are more prone to be biased with respect to the underlying cosmology, as evident from~\Cref{fig:triangle_cal} given our choice of calibration. It is also interesting to note that $f/b$ and $\varepsilon$ are less correlated in the calibrated analysis, since their partial degeneracy with the nuisance parameter $\mathcal{M}$ is removed.

We summarize all of our results in~\Cref{tab:constraints}. The constraints on $f\sigma_8$ and $\DA H$ are derived from the posteriors on $f/b$ and~$\varepsilon$. In the former case, we assume $\xi(r)\propto b\sigma_8$ and, hence, we multiply $f/b$ by the underlying value of $b\sigma_8$ in the Flagship mock, which also assumes the relative precision on $f/b$ and $f\sigma_8$ to be the same. Moreover, we neglect the dependence on $h$ that enters in the definition of $\sigma_8$ and should be marginalized over~\citep{Sanchez2020}. For the latter case, we multiply $\varepsilon$ by the fiducial $\DA H$, following~\Cref{epsilon}. The results on $f\sigma_8$ and $\DA H$ from both model-independent and calibrated analyses are also shown in~\Cref{fig:fs8_DAH}.

\begin{figure*}[t]
        \centering
        \resizebox{\hsize}{!}{
            \includegraphics[scale=0.51, trim=40 8 0 30]{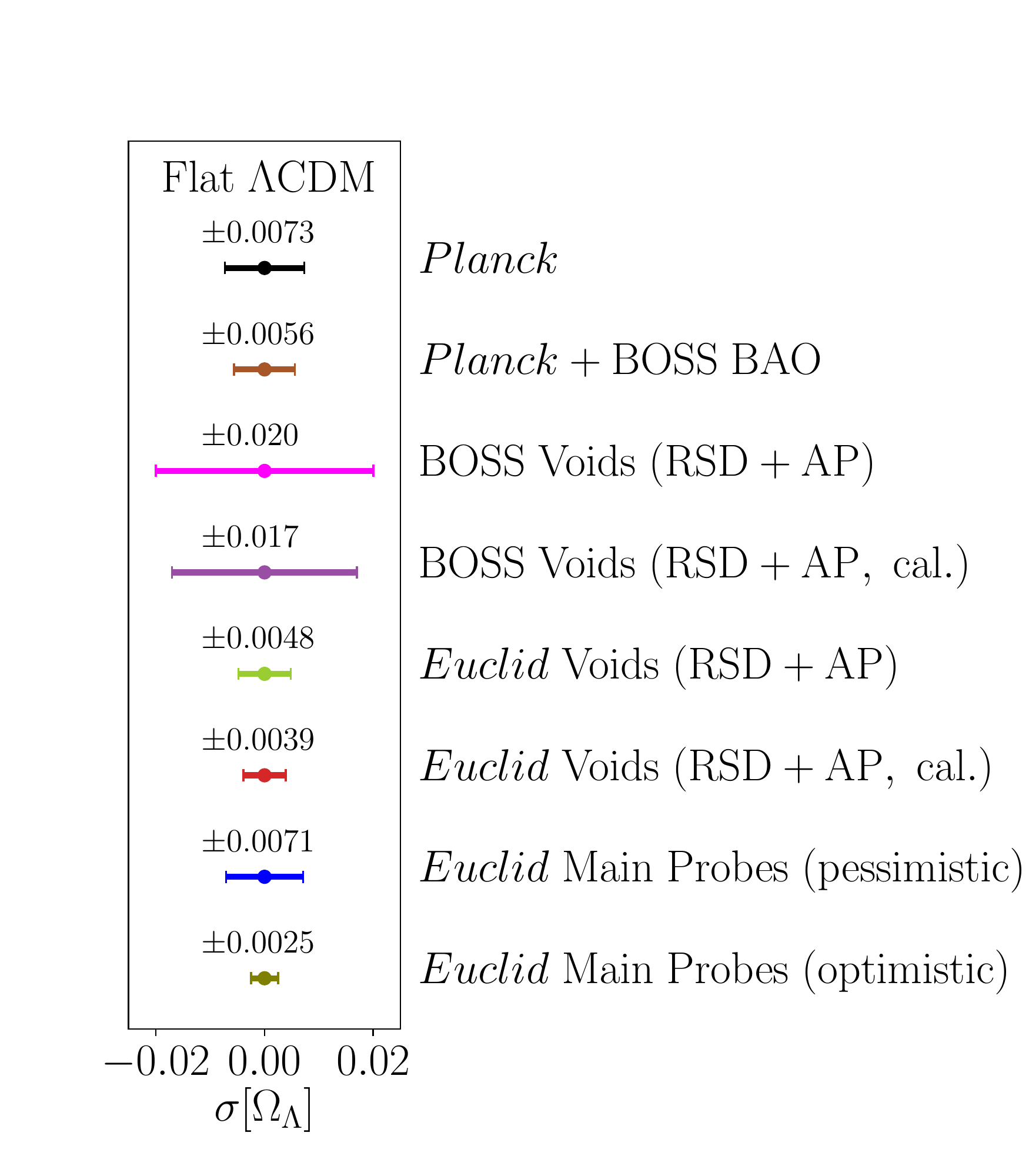}
                \includegraphics[trim=0 5 0 0]{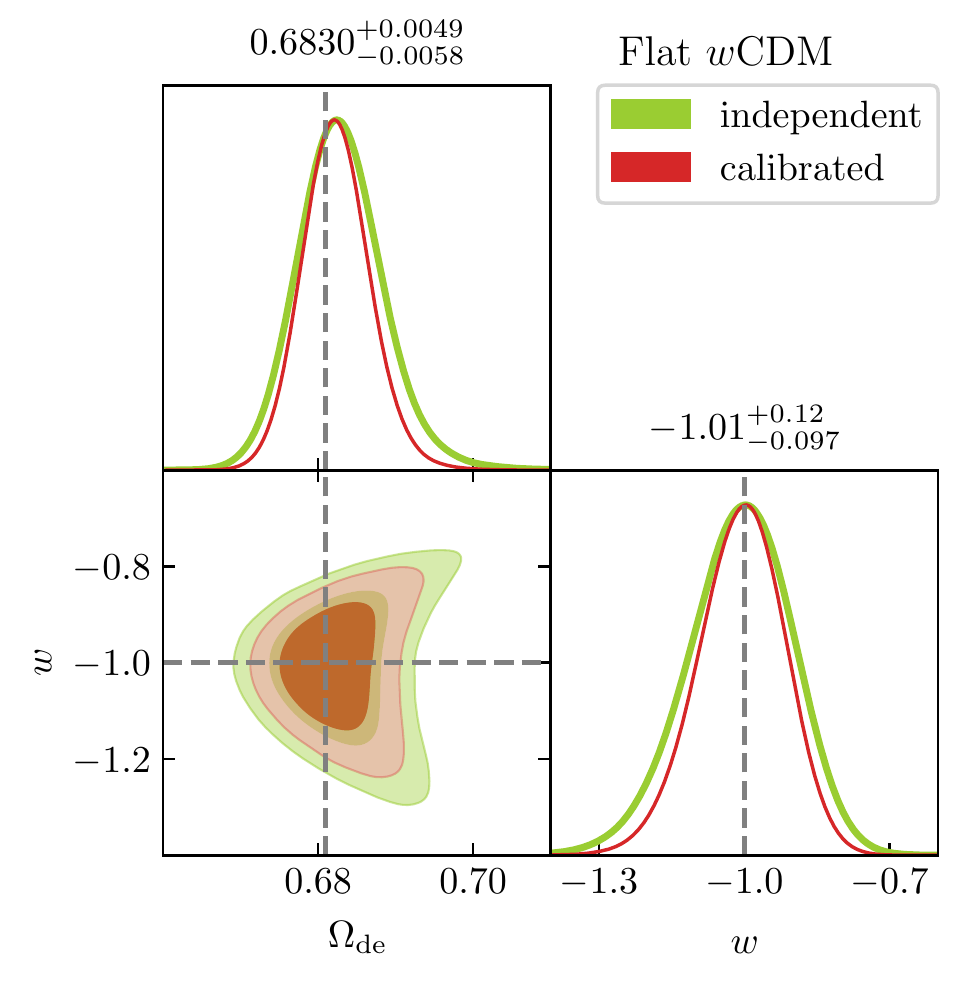}}
        \caption{Forecast for cosmological constraints on dark energy parameters. \textit{Left}: Comparison of the constraining power on $\Ol$ in a flat \lcdm\ cosmology from \Planck 2018 alone~\citep{Planck2018} and when combined with BOSS BAO data~\citep{Alam2017}. Below, constraints as obtained with BOSS voids via RSD and AP~\citep{Hamaus2020}, as expected from \Euclid voids (this work), and as expected from \Euclid's main cosmological probes combined~\citep{Euclid2020}. Calibrated constraints from voids are indicated by the abbreviation ``cal.'' and the \Euclid main probes distinguish between ``optimistic'' and ``pessimistic'' scenarios. \textit{Right}: Predicted constraints from \Euclid voids on dark energy content $\Ode$ and its equation-of-state parameter $w$ in a flat $w$CDM cosmology. Both the model-independent (green) and the calibrated results (red) are shown, dashed lines indicate the input cosmology. Mean parameter values and their $68\%$ confidence intervals for the model-independent case are shown at the top of each panel.}
        \label{fig:cosmology}
\end{figure*}

\section{Discussion}\label{sec:discussion}
The measurements of $f\sigma_8$ and $\DA H$ as a function of redshift can be used to constrain cosmological models. For example, an inversion of~\Cref{HDA} provides $\Ol=1-\Om$, the only free parameter of the product $\DA H$ in a flat \lcdm\ cosmology. Using our Flagship mock measurements of $\DA H$ we sample the joint posterior on $\Ol$ from all of our redshift bins combined. Considering the full \Euclid footprint to be approximately three times the size of our Flagship mock catalog, we scale the errors on $\DA H$ by a factor of $1/\!\sqrt{3}$ and center its mean values to the input cosmology of Flagship. The resulting posterior yields $\Ol=0.6809\pm0.0048$, in the model-independent, and $\Ol=0.6810\pm0.0039$, in the calibrated case from the analysis of \Euclid voids alone. The corresponding result obtained by Planck in 2018~\citep{Planck2018} is $\Ol=0.6847\pm0.0073$, including CMB lensing and $\Ol=0.6889\pm0.0056$, when combined with BOSS BAO data~\citep{Alam2017}. The main cosmological probes of \Euclid, when altogether combined, are forecasted to achieve a $1\sigma$ uncertainty of $0.0071$ on $\Ol$ in a pessimistic scenario and $0.0025$ in an optimistic case~\citep{Euclid2020}. The expected precision on $\Ol$ from the analysis of \Euclid voids alone will hence likely match the level of precision from \Planck and the combined main \Euclid probes. The left panel of~\Cref{fig:cosmology} provides a visual comparison of the constraining power on $\Ol$ from the mentioned experiments, including the one previously obtained from BOSS voids in~\citet{Hamaus2020}.

Furthermore, in \Euclid we aim to explore cosmological models beyond \lcdm. One minimal extension is to replace the cosmological constant, $\Lambda$, by a more general form of dark energy with density $\Ode$, and a constant equation-of-state parameter, $w$. This modifies the Hubble function in~\Cref{HDA} to
\begin{equation}
        H(z) = H_0\sqrt{(1-\Ode)(1+z)^3+\Ode(1+z)^{3(1+w)}}\;,\label{wCDM}
\end{equation}
which reduces to the case of flat \lcdm\ for $w=-1$. Using our rescaled and recentered mock measurements of $\DA H$, we can thus infer the posterior distribution of the parameter pair $(\Ode,w)$. The result is shown in the right panel of~\Cref{fig:cosmology} for both the model-independent and the calibrated analysis, assuming flat priors with $\Ode\in\brackets{0,1}$ and $w\in\brackets{-10,10}$. We observe a mild degeneracy between $\Ode$ and $w$, which can be mitigated in the calibrated case. However, this may come at the price of an increased bias from the true cosmology. Nevertheless, these parameter constraints are extremely competitive, yielding $w=-1.01^{+0.10}_{-0.08}$ and, hence, a relative precision of about $9\%$ in the calibrated scenario. In the model-independent analysis, we still obtain $w=-1.01^{+0.12}_{-0.10}$ with a relative precision of $11\%$. The constraints on $\Ode$ are similar to the ones on $\Ol$ from above. We note that this result is in remarkable agreement with the early Fisher forecast of~\citet{Lavaux2012}, corroborating the robustness of the AP test with voids. The~\cite{Planck2018} obtain $w=-1.57^{+0.50}_{-0.40}$ including CMB lensing in the same $w$CDM model and a combination with BAO yields a similar precision of about $10\%$, with $w=-1.04^{+0.10}_{-0.10}$.

The right panel of \Cref{fig:cosmology} provides a demonstration of how cosmic voids by themselves constrain the properties of dark energy, without the inclusion of external priors, observables, or mock data. A combination with other probes, such as void abundance~(\textcolor{blue}{Contarini et al. in prep.}), cluster abundance~\citep{Sahlen2019}, BAO~\citep{Nadathur2019}, CMB, or weak lensing~(\textcolor{blue}{Bonici et al. in prep.}) will, of course, provide substantial gains. This is particularly relevant when extended cosmological models are considered to enable the breaking of parameter degeneracies. For example, it concerns models with a redshift-dependent equation of state $w(z)$, nonzero curvature, or massive neutrinos. In those cases, however, the calibrated approach is prone to be biased towards the fiducial cosmology adopted in the synthetic mock data used for calibration, which commonly assumes a flat \lcdm\ model. In order to avoid the emergence of confirmation bias, we advocate the more conservative model-independent approach, even if it comes at the price of a somewhat reduced statistical precision.

In our analysis, we have neglected observational systematics that are expected to arise in \Euclid data, such as spectral line misidentification due to interlopers that can lead to catastrophic redshift errors~\citep{Pullen2016,Leung2017,Massara2021}. However, this effect mainly impacts the amplitude of multipoles~\citep{Addison2019,GrasshornGebhardt2019}, so we anticipate it to be at least partially captured by the nuisance parameters in our model. We leave a more detailed investigation on the impact of survey systematics to a future work.

\section{Conclusion}\label{sec:conclusion}
In this work, we investigate the prospects for performing a cosmological analysis using voids extracted from the spectroscopic galaxy sample of the Euclid Survey. The method we applied is based on the observable distortions of average void shapes via RSD and the AP effect. Our forecast relies on one light cone octant ($5157$ $\mathrm{deg}^2$) of the Flagship simulation covering a redshift range of $0.9<z<1.8$, which provides the most realistic mock galaxy catalog available for this purpose to date. Exploiting a deprojection technique and assuming linear mass conservation allows us to accurately model the anisotropic void-galaxy cross-correlation function in redshift space. We explore the likelihood of the mock data given this model via MCMCs and obtain the posterior distributions for our model parameters: the ratio of growth rate and bias $f/b$, and the geometric AP distortion $\varepsilon$. Two additional nuisance parameters, $\mathcal{M}$ and $\mathcal{Q}$, are used to account for systematic effects in the data; they can either be marginalized over (model-independent approach) or calibrated via external sources, such as survey mocks (calibrated approach). After the conversion of our model parameters to the combinations $f\sigma_8$ and $\DA H$, we forecast the attainable precision of their measurement with voids in \Euclid.

We expect a relative precision of about $4\%$ ($1\%$) on $f\sigma_8$ and $0.5\%$ ($0.4\%$) on $\DA H$ without (with) model calibration for each of our four redshift bins. This level of precision will enable very competitive constraints on cosmological parameters. For example, it yields a $0.7\%$ ($0.6\%$) constraint on $\Ol$ in a flat \lcdm\ cosmology and a $11\%$ ($9\%$) constraint on the equation-of-state parameter $w$ for dark energy in $w$CDM from the AP test with voids alone. A combination with other void statistics, or the main cosmological probes of \Euclid, such as galaxy clustering and weak lensing, will enable considerable improvements in accuracy and allow for the exploration of a broader range of extended cosmological models with a larger scope of parameters.
\vspace{20pt}

\begin{acknowledgements}
NH, GP and JW are supported by the Excellence Cluster ORIGINS, which is funded by the Deutsche Forschungsgemeinschaft (DFG, German Research Foundation) under Germany's Excellence Strategy -- EXC-2094 -- 390783311. MA, MCC and SE are supported by the eBOSS ANR grant (under contract ANR-16-CE31-0021) of the French National Research Agency, the OCEVU LABEX (Grant No. ANR-11-LABX-0060) and the A*MIDEX project (Grant No. ANR-11-IDEX-0001-02) funded by the Investissements d'Avenir French government program, and by CNES, the French National Space Agency. AP is supported by NASA ROSES grant 12-EUCLID12-0004, and NASA grant 15-WFIRST15-0008 to the Nancy Grace Roman Space Telescope Science Investigation Team ``Cosmology with the High Latitude Survey''. GL is supported by the ANR BIG4 project, grant ANR-16-CE23-0002 of the French Agence Nationale de la Recherche. PN is funded by the Centre National d'Etudes Spatiales (CNES). We acknowledge use of the Python libraries \texttt{NumPy}~\citep{Harris2020}, \texttt{SciPy}~\citep{Virtanen2020}, \texttt{Matplotlib}~\citep{Hunter2007}, \texttt{Astropy}~\citep{Astropy2013,Astropy2018}, \texttt{emcee}~\citep{Foreman-Mackey2019}, \texttt{GetDist}~\citep{Lewis2019}, \texttt{healpy}~\citep{Gorski2005,Zonca2019}, and \texttt{PyAbel}~\citep{Hickstein2019}. This work has made use of CosmoHub~\citep{Carretero2017,Tallada2020}. CosmoHub has been developed by the Port d'Informació Científica (PIC), maintained through a collaboration of the Institut de Física d'Altes Energies (IFAE) and the Centro de Investigaciones Energéticas, Medioambientales y Tecnológicas (CIEMAT) and the Institute of Space Sciences (CSIC \& IEEC), and was partially funded by the "Plan Estatal de Investigación Científica y Técnica y de Innovación" program of the Spanish government.

\AckEC
\end{acknowledgements}

%\normalem
\bibliographystyle{aa}
\bibliography{ms}

\end{document}